\documentclass[acmsmall]{acmart}
\AtBeginDocument{%
  }

\acmSubmissionID{xxx-xxx-xxx}

\usepackage{multirow}
\usepackage{makecell}
\usepackage[most]{tcolorbox}
\usepackage{bbm}
\usepackage{textcomp}
\usepackage{setspace}
\usepackage{enumitem}

\begin{document}

% \input{Cover/Cover_AE}
% \input{Cover/Cover_R1}
% \input{Cover/Cover_R2}

% \newpage

% \setcounter{page}{1}

\title{Understanding Internal Representations of Recommendation Models with Sparse Autoencoders}

\newcommand\blfootnote[1]{%
    \begingroup
    \renewcommand\thefootnote{}\footnote{#1}%
    \addtocounter{footnote}{-1}%
    \endgroup
}

\author{Jiayin Wang}
\email{JiayinWangTHU@gmail.com}
\orcid{0000-0001-8875-1850}
\affiliation{%
  \institution{Tsinghua University, China. Quan Cheng Laboratory}
  \city{Jinan}
  \country{China}}

\author{Xiaoyu Zhang}
% \email{xiaoyu-z22@mails.tsinghua.edu.cn}
\affiliation{%
  \institution{Tsinghua University}
  \city{Beijing}
  \country{China}}

\author{Weizhi Ma}
% \email{mawz@tsinghua.edu.cn}
\affiliation{%
  \institution{Tsinghua University}
  \city{Beijing}
  \country{China}}

\author{Zhiqiang Guo}
% \email{georgeguo.gzq.cn@gmail.com}
\affiliation{%
  \institution{Tsinghua University}
  \city{Beijing}
  \country{China}}

\author{Min Zhang$^{*}$}
% \email{z-m@tsinghua.edu.cn}
\affiliation{%
  \institution{Tsinghua University, China. Quan Cheng Laboratory}
  \country{China}}
  
% \affiliation{%

% \affiliation{%

% \affiliation{%

% \affiliation{%

% \affiliation{%

\renewcommand{\shortauthors}{Wang et al.}

\begin{abstract}
\blfootnote{*Corresponding Authors}
Recommendation model interpretation aims to reveal the relationships between inputs, model internal representations and outputs to enhance the transparency, interpretability, and trustworthiness of recommendation systems.
However, the inherent complexity and opacity of deep learning models pose challenges for model-level interpretation. Moreover, most existing methods for interpreting recommendation models are tailored to specific architectures or model types, limiting their generalizability across different types of recommenders.

In this paper, we propose RecSAE, a generalizable probing framework that interprets \textbf{Rec}ommendation models with \textbf{S}parse \textbf{A}uto\textbf{E}ncoders.
The framework extracts interpretable latents from the internal representations of recommendation models, and links them to semantic concepts for interpretations. It does not alter original models during interpretations and also enables targeted tuning to models.
Experiments on three types of recommendation models (general, graph-based, sequential) with \textcolor{black}{four widely used public datasets} demonstrate the effectiveness and generalization of RecSAE framework. The interpreted concepts are further validated by human experts, showing strong alignment with human perception. Overall, RecSAE serves as a novel step in both model-level interpretations to various types of recommendation models without affecting their functions and offering potential for targeted tuning of models.
The code and data are available\footnote{https://github.com/Alice1998/RecSAE}.

\end{abstract}







% \received{20 July 2025}

\maketitle

\section{Introduction}
\label{section:introduction}
Transparency is critical in recommendation systems, as it fosters trust, fairness, and accuracy~\cite{zhang2014explicit,du2024tutorial,sinha2002role,ge2024survey}.
Research on recommendation transparency is typically classified into two categories: result-level explanations~\cite{nilashi2016recommendation} and model-level interpretations~\cite{zhang2020explainable}.
Result-level explanations aim to provide end users with reasons for recommendations~\cite{Kaffes2021Conterfactual,ChatRecGao2023}, by revealing relationships between input information and output items.
% focus on explaining the relationships between inputs and outputs~\cite{Kaffes2021Conterfactual,ChatRecGao2023}, providing users with recommendation explanations. 
Model-level interpretations aim to uncover the internal decision‐making rationale of recommendation models~\cite{featureinteractiontsang2020,ClusteringDu2023}, enabling system designers to understand learned representations and further apply targeted tuning to improve model behavior.

\begin{table}[htbp]
\caption{Existing studies on recommendation transparency do not provide model-level interpretations across diverse recommender types, and support targeted tuning of models.}
\label{table:related_work}
 \resizebox{0.95\linewidth}{!}{
  \begin{tabular}{llclc}
    \toprule
     \textbf{Method} & \textbf{Representative Studies} & \textbf{Model-level} & \textbf{General Framework} & \textbf{Targeted Tuning}\\
    \midrule
     Path & 
    Wang 18~\cite{Wang2018KPRN};
    Zhao 20~\cite{KGZhao2020}  & $\sqrt{}$ & &  \\
    % ; Xian 20~\cite{Xian2020CAFE}
 Textual & Hada 21~\cite{ReXPlugHada2021}; Li 23~\cite{Li2023PEPLER} & $\sqrt{}$ & &  \\
 % Li 21~\cite{Li2021PETER}; 
 Attention & Xiao 17~\cite{Xiao2017AFM}; Tal 21~\cite{Tal21Attention} & $\sqrt{}$ &  & \\
 Influence & Liu 19~\cite{In2RecCIKM2019}  &$\sqrt{}$ &  & \\
    Counterfactual & Tan 21~\cite{Tan2021Counterfactual}  & $\sqrt{}$ &  & \\
    
    & Kaffes 21~\cite{Kaffes2021Conterfactual} & &$\sqrt{}$  & \\
    Rule & Zhang 24~\cite{FENCRZhang2024} & $\sqrt{}$ &  & \\
     RL & Wang 18~\cite{RLWang2018} & &$\sqrt{}$ &  \\
    
 LLM & Gao 23~\cite{ChatRecGao2023}; Lei 23~\cite{lei2023recexplainer} & &$\sqrt{}$ &  \\
     Distillation & Xu 23~\cite{Xu2023Distillation}  & & $\sqrt{}$ &  \\ 
     & Zhang 20~\cite{DistillationZhang2020}  & $\sqrt{}$ &$\sqrt{}\mkern-9mu{\smallsetminus}$~*Jointly trained  & \\
    Feature & Tsang 20~\cite{featureinteractiontsang2020}   &$\sqrt{}$ & $\sqrt{}\mkern-9mu{\smallsetminus}$~*Context-based&  \\
    Clustering &  Du 23~\cite{ClusteringDu2023}  & $\sqrt{}$ & $\sqrt{}$ &  \\
    % $\sqrt{}\mkern-9mu{\smallsetminus}$ \checkmark
    \midrule
    Probing & RecSAE~(Ours)  & $\sqrt{}$& $\sqrt{}$ &$\sqrt{}$ \\
  \bottomrule
\end{tabular}
}
% \vspace{0mm}
\end{table}

While model-level interpretations offer notable advantages, interpreting current recommendation models remains challenging due to the opaque nature of neuron activations in deep learning models~\cite{zhang2021survey,montavon2018methods} and the diversity of recommendation model types (such as general, graph-based, context-based, sequential models)~\cite{patel2023recommendation}.
Most existing model-level interpretations rely on specialized model architectures or types~\cite{Wang2018KPRN,ReXPlugHada2021,Xiao2017AFM,ClusteringDu2023,FENCRZhang2024,balog2019transparent}, limiting their applicability to different types of recommendation models. As shown in Table~\ref{table:related_work}, three studies aim to develop more generalized interpretation methods. Tsang et al.~\cite{featureinteractiontsang2020} introduced a feature interaction interpretation method for context-based recommendation models.
Another distillation-based approach~\cite{DistillationZhang2020} learns structural knowledge from an explainable model to embedding-based models. However, it requires joint training with the target model, reducing its adaptability to pre-existing models. Du et al.~\cite{ClusteringDu2023} propose a clustering-based method, learning interpretable clusters with interaction logs and item tags. However, it sacrifices performance and necessitates model ensemble to compensate. Moreover, the above methods cannot apply targeted modifications to model behaviors. To the best of our knowledge, no existing work establishes the capability for interpreting diverse types of recommendation models without affecting their performance and simultaneously enabling targeted tuning to models.

Motivated by recent advances in employing sparse autoencoders (SAEs) to interpret transformer-based large language models~\cite{claude_towards, claude_scaling, gao2024scaling}, we utilize the SAE structures to the interpretation of recommenders.
SAEs employ large-dimensional, sparsely activated latents to decompose poly-semantic model representations into mono-semantic latents through unsupervised learning objectives, thereby extracting interpretable concepts embedded in the original model. 
\textcolor{black}{Besides, sparse autoencoders are suitable for recommendation tasks. User–item interactions in recommender systems are inherently sparse, as most items remain unseen by a user. Meanwhile, user interests and profiles are concept- or feature-level rather than item-level~\cite{godoy2006modeling}. While item-level preferences are explicitly observable through user interactions (e.g., clicks or purchases), the underlying concept- or feature-level preferences are latent and not directly reflected in user logs. Therefore, if the model can effectively extract core concept-level representations from sparse item-level data, it would enhance both the accuracy and interpretability of user modeling, as well as the accuracy and interpretability of the resulting recommendations.}

Beyond the potential of SAEs in recommendation scenarios, it is worth noting that existing SAE interpretation pipelines have primarily been validated in the context of large language models. However, recommendation models differ fundamentally from LLMs in terms of architecture, training data, and optimization objectives, as further discussed in Section~\ref{subsection:sae}.
Various types of recommenders, including general and graph-based models, do not incorporate transformer structures.
However, the effectiveness of SAEs for interpretation has been demonstrated primarily within specific structures in transformers.
% Thus, it still need further designs and validation of adopting SAEs to interpret various types of recommendations.
Consequently, unlocking the potential of SAEs for recommendation model interpretations demands design adaptations and rigorous validation across diverse types of recommendation models.

In this work, we introduce an effective probing paradigm for interpreting various types of recommendation models without altering their functionalities.
% In this paper, we propose the probing paradigm for interpreting various kinds of recommendation models. 
% It enables model interpretation without altering their functionality or performance.
It also enables target tuning to models.
% The proposed RecSAE framework extracts interpretable latents from the internal states of existing recommenders and associates these latents with semantic concepts for model interpretation.
RecSAE interpretation operates in three steps:
(1) \textbf{Probe Internal Representations}: RecSAE first probes the internal user representations in recommendation models.
(2) \textbf{Train RecSAE Module}: The RecSAE module is trained on these representations with a larger latent space and under sparsity constraints, making RecSAE latents more interpretable than the original model representations.
(3) \textbf{Construct and Verify Semantic Concepts}: 
RecSAE constructs concept descriptions using three sources through two independent methods.
The three sources are input items, RecSAE latent activations, and recommendation‐model outputs.
The two methods are TF–IDF algorithm (term frequency–inverse document frequency) to derive term‐level interpretations, and a large language model (Llama-3-8b-Instruct~\cite{llama3modelcard}) to generate semantic‐level explanations.
%RecSAE evaluates using two method to construct concept descriptions based on input items, latent activations and recommendation model outputs. The two methods are the TF-IDF algorithm (term frequency–inverse document frequency) for term-level interpretation and a language model (Llama-3-8b-Instruct~\cite{llama3modelcard}) for sematic-level interpretation. 
These interpreted concepts are validated by predicting latent activations on new cases and calculating confidence scores based on prediction accuracy. Human experts further verify the constructed concepts and reach an average relevance score of 3.89/4, showing the efficacy of the automated concept construction pipeline. Additionally, RecSAE enables targeted tuning of the interpreted models by replacing the original representations at the probing position with the reconstructed representations.
In summary, our contributions are threefold:

\begin{itemize}
    \item We propose the RecSAE framework with a probing paradigm for interpreting various types of recommendation models. This pipeline enables model interpretation without affecting the performance of well-trained models.

    % \item RecSAE enables model interpretation without affecting the performance of well-trained models, \todo{while also allowing targeted tuning of models based on interpretation results.}
    \item RecSAE extracts interpretable latents from the internal states of recommendation models and links them to semantic concepts with interpretation confidence scores. These concepts are human-readable and easy to verify.

    \item Experiments on three types of recommendation models — general, graph-based, and sequential — across widely used datasets demonstrate the effectiveness and generalizability of the RecSAE framework. These results highlight RecSAE's capability to provide model-level interpretations and its potential to facilitate targeted tuning of models.

\end{itemize}

\section{Related Work}
\label{section:related_work}

\subsection{Explainable and Interpretable Recommendation}
\label{subsection:related_work_rec}

Table~\ref{table:related_work} summarizes studies on interpretable and explainable recommendation systems, categorizing methods based on the interpreted subject (result-level or model-level), their generalization (specific to a particular model, a type of model, or across various models), and their ability to enable targeted tuning of the interpreted models.

Result-level interpretation approaches, such as explanation distillation~\cite{Xu2023Distillation}, LLM augmentation~\cite{ChatRecGao2023}, and reinforcement learning~\cite{RLWang2018}, offer generalizability but do not fully reveal the mechanisms of models.
In contrast, model-level interpretation methods generate insights into model structures using techniques like path-reasoning~\cite{Wang2018KPRN,KGZhao2020,Xian2020CAFE}, textual modeling~\cite{ReXPlugHada2021,Li2021PETER,Li2023PEPLER}, attention mechanisms~\cite{Xiao2017AFM,Tal21Attention}, counterfactual methods~\cite{Tan2021Counterfactual}, and clustering~\cite{ClusteringDu2023}. However, these methods are often constrained by dependence on specific model architectures, limiting their generalization.
Beyond them, three studies offer a certain degree of generalization. 
Zhang et al.~\cite{DistillationZhang2020} transferring structural knowledge from a path-based explainable model. However, as this framework requires target interpretation models to be trained jointly with the explainable model, it limits its adaptability to pre-existing models.
Tsang et al.~\cite{featureinteractiontsang2020} propose a feature-based interpretation method, but the approach is tailored to context-based recommendation models utilizing feature interactions. 
Du et al.~\cite{ClusteringDu2023} mine taste clusters from user-item interactions and item tags. However, their clustering paradigm impacts model diversity and performance, requiring ensemble learning with at least three models, thereby increasing training costs.
\textcolor{black}{
With the development of language-model-based generative recommendation, studies have explored learning semantic IDs to represent items in a discrete, language-like form~\cite{rajput2023recommender, wang2024learnable}.
These works reveal a misalignment between semantic identifiers and collaborative signals.
In contrast, our work focuses on interpreting traditional recommendation models that are purely trained on collaborative information, aiming to understand the semantics implicitly encoded in their latent representations.}

In summary, existing model-level interpretation methods exhibit limited generalization capabilities to various kinds of recommendation models, and to the best of our knowledge, no related paradigm could utilize interpreted results for targeted model debiasing.

\subsection{Sparse Autoencoder}
\label{subsection:sae}

Interpreting deep learning models to understand their decision-making processes is a critical research area~\cite{liang2021explaining, jazayeri2021interpreting,sajjad2022neuron, foote2023neuron, fan2024evaluating}. The field of neuron interpretation has recently advanced through several related work, including interpreting single-layer transformers~\cite{claude_towards} and analysis of large language models such as GPT-4 and Claude-3~\cite{gao2024scaling, claude_scaling}.
Specifically, Anthropic has employed dictionary learning~\cite{tovsic2011dictionary} to interpret neuron activations in single-layer transformers. Additionally, several studies have analyzed residual stream activations in transformers using sparse autoencoders~\cite{cunningham2023sparse, claude_scaling, gao2024scaling}, aiming to uncover mono-semantic features and improve the interpretability of language models. OpenAI has also leveraged language models to interpret neuron activations at scale~\cite{openai_exp_neuron,gao2024scaling}.
\textcolor{black}{Though sparse autoencoders show great potential for interpreting deep learning models, they also face several inherent limitations. In practice, issues such as activation shuffling, over-sparsification, and the lack of guaranteed clean or mono-semantic features remain~\cite{sae_limitation_anthropic, sae_limitation_iclr}. Consequently, their effectiveness in interpreting diverse recommendation models trained on inherently sparse user–item interaction data remains largely unexplored.}

\textbf{Besides, recommendation systems exhibit several fundamental differences}.
Firstly, not all recommendation models are transformer-based. A wide variety of models—such as general, sequential, graph-based, and context-based, content-based models—exist, each with distinct structural characteristics. While sparse autoencoders have demonstrated effectiveness primarily on residual activations within transformer structures~\cite{claude_scaling, gao2024scaling}, their applicability to other architectures remains underexplored. Additionally, recommendation models are trained on interaction logs to learn collaborative signals, where ID-based training data lack direct correspondence to semantic concepts. Third, the evaluation of interpretability in recommendation systems differs from that in language models, as items are often associated with explicit attributes, and prior work~\cite{ClusteringDu2023} employed intra- and inter-attribute similarity for assessment. 
\textcolor{black}{Despite these differences, SAEs hold strong potential in recommendation scenarios.
Recommendation models operate on high-dimensional and sparse interaction spaces, while user interests are typically expressed at the feature level rather than the item level.
By leveraging sparse autoencoder techniques, user–item co-occurrence patterns can be effectively disentangled into interpretable latent factors aligned with human-understandable concepts.
While some prior studies have incorporated SAE-based structures for recommendation tasks (e.g., SLIM~\cite{ning2011slim} and EASE~\cite{steck2019embarrassingly}), RecSAE differs fundamentally in purpose and scope. These earlier methods aim to build standalone models for improving recommendation accuracy, whereas RecSAE functions as a general interpretation framework—designed to probe and explain existing recommendation models through semantically meaningful latent concepts, thereby enabling both interpretability and targeted tuning.}
In this paper, we introduce the RecSAE framework, which is designed to interpret various ranges of recommendation models with semantic concepts and provides similarity- and confidence-based scores to quantify the reliability of these interpretations.

\section{RecSAE}
\label{sec:method}
\subsection{Overview}
\label{subsection:overview}

Formally, the interpretation task is defined as follows.
Given 1) a recommendation model, $M_{rec}$; 2) ID-based interaction logs $L = \{(ID_{u}, ID_{i}), ...\}$~(content-based interaction log are also applicable within the framework); 3) item descriptions $D = \{d_{i}, ...\}$, such as item titles or/and categories, the objective is to use unsupervised learning method to extract mono-semantic latents from the model's internal representations and interpret them into human-understandable concepts. This process aims to reveal the correspondence between the model’s internal representations and interpretable semantic concepts.

\begin{figure*}[htbp]
    \centering
     \includegraphics[width=\textwidth]{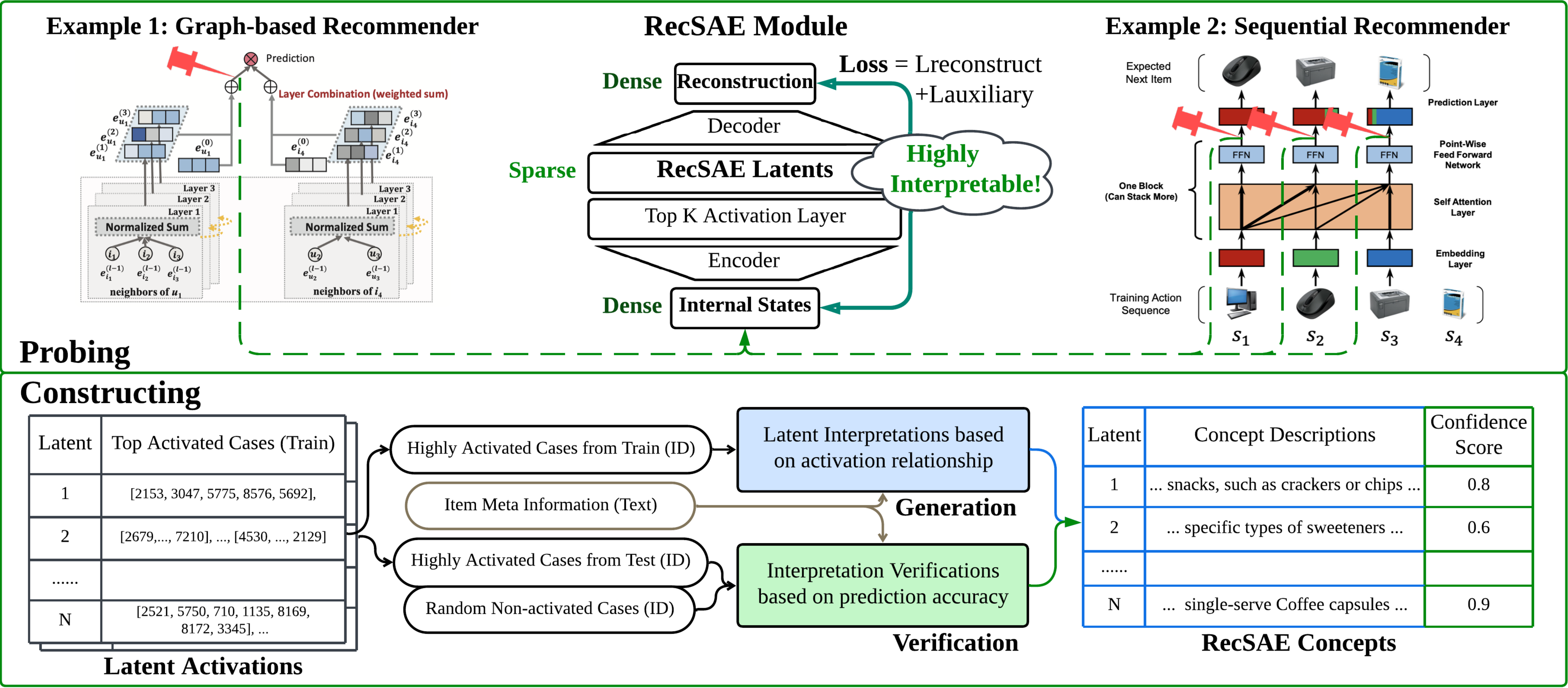}
    \caption{\textcolor{black}{RecSAE Framework.} It encodes the inner representations of recommendation models into large and sparse latent space and constructs corresponding concepts with confidence scores based on the relationship between recommendation input, latent activations, and recommendation outputs. Example recommendation models are LightGCN~\cite{he2020lightgcn} and SASRec~\cite{kang2018self}.}
    \label{fig:framework}
\end{figure*}

Based on this general setting, we propose RecSAE, a probing framework that uses a sparse autoencoder structure to interpret the internal states of recommendation models. 
\textcolor{black}{As illustrated in the Introduction, sparse autoencoder techniques are well-suited to recommendation scenarios. User–item interactions are inherently sparse, as most items remain unseen by a given user. However, user interests are typically expressed at the feature or concept level rather than the item level. This motivates concept-level modeling to better interpret user preferences. Sparse autoencoder techniques are well-suited to these scenarios.}
By encoding model representations into a large, sparse latent space, RecSAE uncovers textual concepts that shape recommendation outcomes, providing insights into previously opaque models.
As illustrated in Figure~\ref{fig:framework}, RecSAE features a probing architecture and an automated interpretation pipeline, generating descriptions for each latent and validating these interpretations with confidence scores.
The following subsection~\ref{subsec:model_architecture} and \ref{subsec:auto_interpretation} provide detailed descriptions.

\subsection{RecSAE Training}
\label{subsec:model_architecture}

In this section, we describe the unsupervised learning process of RecSAE, which extracts interpretable and mono-semantic latents from the internal representations of recommendation models. To promote interpretability, RecSAE employs a large latent dimensionality combined with sparsity-controlled activation mechanism~\cite{gao2024scaling}. This design disentangles the original dense, poly-semantic representations into a larger number of sparse, mono-semantic latents.

In the following sections, we introduce RecSAE's probing site, module architecture, and the associated training loss function.

\subsubsection{\textbf{Probing Site}}

User‐interest representations encode the personalized information learned by recommendation models~\cite{ko2022survey}. These representations are typically formed before the prediction layer for similarity computations with candidate items. Accordingly, as illustrated in Figure~\ref{fig:framework}, we probe at the user representations prior to the prediction layer. Future work could investigate additional probing sites to gain deeper insights. RecSAE is trained to reconstruct the representations extracted at the probing site.
Formally,

\begin{equation}
\begin{split}
   M_{rec}(u) &= i_{pred} \\
   M_{recP1}(u) = S_{Internal}, \ & M_{recP2}(S_{Internal}) = i_{pred} \\
\end{split}
\end{equation}

where
$M_{rec}$ represents the interpreted recommendation model,
$u$ denotes the user identifier or interaction history,
$i_{pred}$ represents the model's predicted item, 
$S_{Internal}$ is the user‐interest representation at the probing site,
$M_{recP1}$ and $M_{recP2}$ denotes  the two sub‐networks before and after the probing site.

% In recommendation models, user interest representations capture the personalized information encoded by the model~\cite{ko2022survey}.
% These representations are typically formed at the final layer before similarity computation with candidate items. Therefore, as shown in Figure~\ref{fig:framework}, we probe at these user representations prior to the prediction layer in this paper. As discussed in Section~\ref{sec:discussion}, future work could explore probing at different sites to gain deeper insights. RecSAE is trained to reconstruct the representations at the probing site.

\subsubsection{\textbf{Module Architecture}} 

Unlike standard autoencoders, RecSAE scales the input by a factor of $s$ relative to the original representation dimension and incorporates an additional sparsity objective to promote mono-semanticity~\cite{claude_scaling} in the latent space. A vanilla approach to achieve sparsity is to apply an $L_1$ regularization term to the training loss. However, $L_1$ is an imperfect approximation of $L_0$ and it introduces a bias of shrinking all positive activations toward zero~\cite{gao2024scaling}.

% in a dense distribution of low-activation latents, which does not align to achieve $L_0$ sparsity in the latent space $S_{Latent}$.

% \todo{stop here!}
To address this, RecSAE employs a Top-k activation layer, which explicitly controls the number of active latents after encoding. The Top-K activation layer has been proved to outperforms other activation layers (ReLU, ProLU STE, Gated) on sparsity-reconstruction frontier in sparse autoencoders~\cite{gao2024scaling}. By contrast, using a sparsity‐inducing penalty term requires careful tuning of a regularization weight, which can vary across different recommendation types. Consequently, we adopt the Top-K activation mechanism to ensure stable sparsity control in RecSAE.
The formal module structure is illustrated by the following equations, where $S_{Latent}$ denotes the sparse latent vectors with $k$ latents activated and the others set to 0, and $\hat{S}_{Internal}$ is the reconstructed representations. $M_{Pe}$ and $M_{Pd}$ represent the encoder module and decoder module of the RecSAE module. $W_{enc}$, $b_{pre}$, $W_{dec}$ represent the encoder weight, bias and decoder weight. The sensitivity of RecSAE to its hyper-parameters, scale size $s$ and sparsity factor $k$, is further discussed in Section~\ref{section:recstruct_interpet}.

\begin{equation}
\begin{split}
    % z &= \text{TopK}(W_{enc}(x-b_{pre})) \\
    % \hat{x} &= W_{dec}z + b_{pre} \\
    S_{Latent} &= M_{Pe}(S_{Internal}) = \text{Top-k}(W_{enc}(S_{Internal}-b_{pre}))\\
    \hat{S}_{Internal} &= M_{Pd}(S_{Latent})= W_{dec}S_{Latent} + b_{pre}
\end{split}
\end{equation}

\subsubsection{\textbf{Training Loss}}

RecSAE adopts an unsupervised learning approach, minimizing the $L_2$ distance between $S_{Internal}$ and $\hat{S}_{Internal}$ to train the module. This process ensures that the large, sparse latent space $S_{Latent}$ effectively encodes the information from the internal representations. Thus, interpretations of the latent space reflect information encoded in the interpreted model.

In addition to the primary optimization objective, training RecSAE with Top-k activation introduces a key challenge: preventing "dead" latents, units that remain inactive during training or inference, to fully utilize the expanded latent space. While activating $k$ latents per input improves efficiency and interpretability, it also means that inactive latents receive no gradient updates from the reconstruction loss and therefore are not effectively trained.
To address this, we introduce an auxiliary loss, drawing inspiration from previous work~\cite{gao2024scaling, claude_ghost_grads}. The auxiliary loss targets these inactive latents during training, guiding their learning even though they are not involved in the main reconstruction process.
Formally, given the reconstruction error $e$, the auxiliary loss $\mathcal{L}_{aux}$ is defined as the $L_2$ distance between the main reconstruction error and the reconstruction using top $k_{auxD}$ dead latents, as shown in the equations below.
Since the encoder preceding the Top-k activation layer is shared, adding this auxiliary loss incurs limited computational overhead.
The training loss is the linear combination of $\mathcal{L}$ and $\mathcal{L}_{aux}$. 
% The effectiveness of RecSAE training designs is validated in Section~\ref{subsection:training_effectiveness}.

\begin{equation}
\begin{split}
    \text{loss}  &= \mathcal{L} + \alpha \mathcal{L}_{aux} \\
    \mathcal{L} & = ||S_{Internal}-\hat{S}_{Internal}||^2_2 \\
    \mathcal{L}_{aux}   &= || e-\hat{e} ||^2_2, \quad\text{where } e = S_{Internal}-\hat{S}_{Internal}, \\  \hat{e}& = W_{dec}S_{L}', \  S_{L}' = \text{Top-k}_{auxD}(W_{enc}(S_{Internal}-b_{pre})).
\end{split}
\end{equation}

With the above structure and training strategy, RecSAE is capable of probing various types of recommendation models. While applied to well-trained models, RecSAE allows for interpretation without altering model functions. Moreover, by modifying latent activations and replacing the original representations with their modified reconstructions, RecSAE facilitates targeted tuning of models. This potential is discussed in Section~\ref{section:targeted_debiasing}.

\subsection{RecSAE Concept Construction and Verification}
\label{subsec:auto_interpretation}

After training, RecSAE establishes the mapping relationships between inputs, RecSAE latent activations, and model outputs. Using these links, we use an automated, closed-loop pipeline for generating and validating semantic concepts for each latent.

\textcolor{black}{For interpretation dataset construction, we build an interpretation training set for concept construction and an interpretation test set for concept verification. The recommendation test set is reserved for interpretation verification, and the recommendation training set is used for interpretation construction. This setup ensures that RecSAE concept construction is conducted without data leakage. As illustrated in the lower part of Figure~\ref{fig:framework}, the process first records the input, latent activations, and recommendation outputs obtained during the base recommendation model’s inference on both the training and test sets. Note that for the main experiment results, the recommendation outputs refer to the top-1 recommended item.}
% RecSAE uses the test set interactions for concept construction and verification.
\textcolor{black}{Next, we partition the continuous latent activations into 11 intensity levels (0–10): level 0 denotes non-activation, and levels 1–10 are equal-frequency bins derived from the training-set distribution.}
Then, RecSAE selects the top 2N most activated cases for each latent, randomly assigning N cases for concept construction and reserving the remaining N for verification. Additionally, N non-activated cases for each latent are randomly selected for verification.
In line with OpenAI's setting to interpret neurons with language models~\cite{openai_exp_neuron}, we set N=5 in this study.

For concept construction, RecSAE generates textual descriptions for each latent, transforming latents into textual concepts, as detailed in Subsection~\ref{subsection:feature_interpretation_construction}.
For concept verification, RecSAE evaluates the interpretations by predicting activation levels on new inputs and calculating precision and recall scores to determine interpretation confidence, as outlined in Subsection~\ref{subsection:feature_interpretation_verification}.
We employ two methods in the above two-stage pipeline: (1) the TF-IDF~(term frequency–inverse document frequency) algorithm for term-level statistical summarization; (2) a large language model for semantic-level analysis. The statistical method provides a reliable baseline, while the LLM offers a strong reference point for the upper bound of automatic interpretation performance.

\subsubsection{Construction}
\label{subsection:feature_interpretation_construction}

RecSAE can latents effectively extract items with notable similarity, as further shown in Table~\ref{table:examples}. Consequently, summarizing these similarities into coherent descriptions is a relatively straightforward task. Therefore, we adopt both term-level algorithm and an open-sourced LLM, instead of relying on paid APIs with state-of-the-art close-sourced large language models. We opt the TF-IDF algorithm~\cite{christian2016single} and a locally deployable model, Llama-3-8B-Instruct~\cite{llama3modelcard}, to generate concept descriptions in this paper.
For the TF-IDF algorithm, we construct the textual representation for each latent using the metadata of items in the interpretation set, and identify its concept as the term with the highest TF-IDF score.

For LLM interpretation, the instruction starts with a system prompt outlining the latent interpretation task, followed by a one-shot example that includes a query and the corresponding assistant response of concept descriptions. Then, the interpreted latents' activation cases are provided for interpretation.
The complete instruction with one-shot example for each dataset is presented in our anonymous repository.

\begin{tcolorbox}[title = {Instruction for Concept Construction},breakable]

\small
\textbf{<Role.SYSTEM>}\\
We're studying neurons in a recommendation model that is used to recommend \{item\_category\}. Each neuron looks for some particular concepts in \{item\_category\}. Look at the parts of the \{item\_name\} the neuron activates for and summarize in a single sentence what the neuron is looking for. Don't list examples of words.\\
The activation format is \{item\_name\}<tab>activation. Activation values range from 0 to 10. A neuron finding what it's looking for is represented by a non-zero activation value. The higher the activation value, the stronger the match. \\

\textbf{<Role.USER>}\\
Neuron 0\\
Activations:\\
\{example\_item\_activation\_sequences\}\\
Explanation of neuron 0 behavior: the main thing this neuron does is find\\

\textbf{<Role.ASSISTANT>}\\
 \{example\_explanation\}\\
 
\textbf{<Role.USER>}\\
Neuron \{latent\_id\}\\
Activations:\\
\{item\_activation\_sequences\}\\
Explanation of neuron \{latent\_id\} behavior: the main thing this neuron does is find
\end{tcolorbox}

Specifically, the above instruction analyzes item activation sequences. 
Each sequence includes the user's interaction history, the model's predicted item, and the corresponding activation level on the interpreted latent.
An example of such a sequence is shown below:

\vspace{0.5\baselineskip}
\noindent
\textit{
<start> \\
% KIND Fruit \& Nut Bar, Almond \& Apricot, 1.4-Ounce Bars (Pack of 12)\quad(0) \\
Bob's Red Mill Organic Scottish Oatmeal, 20-Ounce Bags (Pack of 4)\quad(0) \\
Funky Monkey Snacks Carnaval Mix, Freeze-Dried Fruit, 1-Ounce Bags (Pack of 12)\quad(0) \\
Mrs. May's  Trio Bar Variety Pack, 1.2-Ounce bars (Pack of 20)\quad(0) \\
Mrs. May's Strawberry Trio Bar, 1.2-Ounce Bars (Pack of 24)\quad(10) \\
<end>}
\vspace{0.5\baselineskip}

\textcolor{black}{Since the model activates the relevant latent when predicting the recommended item, the activations occur on the predicted item. As shown in the example, a latent activation sequence consists of the user’s historical interacted items, serving as contextual input with zero activation on the latent, and the model’s predicted items, which exhibit the corresponding activation levels on that specific latent.}
With the above instruction and the activated sequence information of each latent, the pipeline use the Llama model to form the interpreted concepts.

\subsubsection{Verification}
\label{subsection:feature_interpretation_verification}

To assess the quality of the automatically generated concepts, RecSAE evaluates them through confidence scores. Specifically, it leverages the concept to predict latent activations on new instances. Specifically, we use two types of test cases: highly activated cases (distinct from those used for concept constructions), and randomly selected cases with no activation. 
\textcolor{black}{If a concept appears in a negative case, it is counted as a failure; if it appears in a positive case, it is counted as a hit. The use of both sets enables the calculation of confidence scores and ensures that the verification process is discriminative.}

For the TF-IDF method, verification is performed through exact term matching. If the extracted concept term appears in the metadata of an item within the verification set, it is considered a hit; otherwise, it is marked as a miss. The interpretation confidence score for each latent is then computed as the proportion of hit cases among all cases in the verification set.

For LLM method, the verification process begins with a system prompt that outlines the activation-level prediction task, followed by a one-shot example. This example includes a user query containing the concept description, a test case, and the assistant’s predicted activation level. The template is available in the codebase.
After predicting latent activation levels, RecSAE computes a confidence score for each concept. A test case is labeled as positive if its predicted activation exceeds 5; otherwise, it is labeled negative. The confidence score is computed as the proportion of correctly classified cases (i.e., true positives and true negatives) among tests, reflecting the accuracy of the concept description in representing latent semantics. 
The formal calculation of the confidence score is given in Equation~\ref{equation:precision}, where $\text{pos}$ and $\text{neg}$ denote the sets of activated and non-activated test cases, respectively; $level(x)$ is the LLM-predicted activation level for test case $x$; $thres$ is the classification threshold; and $\mathbbm{1}$ is the indicator function.

\begin{equation}
\label{equation:precision}
    % S_{\rm{conf}}(l)
    Precision_{concept} = \frac{\sum_{p \in \text{pos}} \mathbbm{1}[level(p)>thres] + \sum_{n \in \text{neg}} \mathbbm{1}[level(n)<=thres]}{|\text{pos}| + |\text{neg}|}
\end{equation}

% \todo{equation explanations.} Items are labeled into positive and negative with the corresponding concepts present in item descriptions....

\begin{tcolorbox}[title = {Instruction for Concept Verification},breakable]
\small

\textbf{<Role.SYSTEM: 'system'>}

"We're studying neurons in a recommendation model that is used to recommend \{item\_category\}. Each neuron looks for some particular concepts in \{item\_category\}. Look at an explanation of what the neuron does, and try to predict its activations on a particular \{item\_name\}.

The activation format is \{item\_name\}<tab>activation, and activations range from 0 to 10. Most activations will be 0.

Now, we're going predict the activation of a new neuron on a single \{item\_name\}, following the same rules as the examples above.  \\
% Activations still range from 0 to 10. \\

\textbf{<Role.USER: 'user'>}

Neuron \{example\_latent\_id\} \\
Explanation of neuron \{example\_latent\_id\} behavior: the main thing this neuron does is find \{example\_explanation\}  \\
Sequence:\\
\{example\_sequence\}

Last product in the sequence:\\
\{example\_last\_item\}

Last product activation, considering the product in the context in which it appeared in the sequence: \\

\textbf{<Role.ASSISTANT: 'assistant'>}

\{example\_last\_item\_activation\}\\

\textbf{<Role.USER: 'user'>}

Neuron \{test\_latent\_id\} \\
Explanation of neuron \{test\_latent\_id\} behavior: the main thing this neuron does is find \{test\_explanation\}  \\
Sequence:\\
\{test\_sequence\}

Last product in the sequence:\\
\{test\_last\_item\}

Last product activation, considering the product in the context in which it appeared in the sequence:

\end{tcolorbox}

In Section~\ref{subsubsection:human_evaluation}, we further validate the effectiveness of RecSAE’s automated concept construction and verification pipeline through expert annotation. Concepts with confidence scores $\geq 0.8$ are selected for human evaluation to assess their quality.

\section{Experimental Setup}
% In this section, we introduce the experimental setup of our experiments.
% Section~\ref{subsection:dataset} introduced the three public available and widely used datasets adopted in our experiments.
% Section~\ref{subsection:models} introduces the three types of recommendation models that we used for interpretation and targeted tuning evaluations.
% Section~\ref{subsection:baseline} introduces the baseline for comparing with RecSAE in interpretation.
% Section~\ref{subsection:metrics} presents the evaluation metrics for both interpretation and RecSAE reconstruction performances.
% Section~\ref{subsection:hyperparameter} represents the hyper-parameters in RecSAE implementations and there are also hyper-parameter sensitivity experiments in later section~\ref{section:recstruct_interpet}.

In this section, we describe the experimental setup used in our study. 
We implement the RecSAE framework based on a publicly available recommendation framework, ReChorus~\cite{li2024rechorus2}, which is recommended by the RecSys Conference\footnote{https://github.com/ACMRecSys/recsys-evaluation-frameworks}.
Section~\ref{subsection:dataset} provides an overview of the \textcolor{black}{four} publicly available and widely used datasets employed in our experiments. Section~\ref{subsection:models} introduces the three types of recommendation models used for interpretations. Section~\ref{subsection:baseline} outlines the baseline models for comparison with RecSAE in terms of interpretation. Section~\ref{subsection:metrics} details the evaluation metrics used to assess both interpretation and RecSAE reconstruction performance. Section~\ref{subsection:hyperparameter} presents the hyper-parameters used in the RecSAE implementation, and their sensitivity experiments is further discussed in experiment Section~\ref{section:recstruct_interpet}.

\subsection{Dataset}
\label{subsection:dataset}
We utilize \textcolor{black}{four} widely used datasets: (1) Amazon~\cite{leskovec2007dynamics}: the Grocery \& Gourmet Food subset of a large-scale Amazon Reviews dataset.
(2) MovieLens 1M~\cite{harper2015movielens}: a rating data set from the MovieLens website.
(3) LastFM~\cite{schedl2022lfm}: listening records of the music platform Last.fm.
\textcolor{black}{(4) Yelp~\cite{Yelp2018}: a dataset of Yelp businesses, reviews, and user data from Yelp Challenge 2018.}
    % \item MIND~\cite{wu2020mind}: a large-scale news recommendation dataset derived from the Microsoft News website.

We adopt the built-in data settings and processing procedures provided by ReChorus.
\textcolor{black}{Specifically, for the Amazon dataset, positive interactions refer to users who have reviewed items; for the MovieLens dataset, positive interactions correspond to ratings equal to or above 4; for the LastFM dataset, positive interactions represent listening events between a user and a track; for the Yelp dataset, positive interactions represent reviews with ratings equal to or above 4.}
For data processing, the framework applies a 5-core filter, and performs a leave-one-out split for training, validation, and testing sets. Table~\ref{tab:dataset_statistics} lists the statistics of \textcolor{black}{four} datasets. Both datasets contain ID-based interactions and item textual descriptions.

\begin{table}[htbp]
\caption{Dataset Statistics}
    \label{tab:dataset_statistics}
    \centering
    \begin{tabular}{lrrr}
    \toprule
    \textbf{Dataset} & \textbf{\#Users} & \textbf{\#Items} & \textbf{\#Interactions}\\ \hline
    % \cmidrule(lr){2-4}
        Amazon & 14,681 &8,713 & 151,254\\
        MovieLens &6,032 &3,125 & 574,197 \\
        % LastFM  & 1,359 & 38,465 & 352,325\\
        LastFM & 8,031 & 27,065 & 256,699 \\
        \textcolor{black}{Yelp} & \textcolor{black}{180,156} & \textcolor{black}{79,601} &  \textcolor{black}{2,559,541} \\
        \bottomrule
        % MIND & 267,742 & 8,572 & 2,765,523 \\
        % "# user": 18637, "# item": 14458, "# entry": 211108

    \end{tabular}

\end{table}

For SAE training, we infer the interpreted recommendation model over the training set and record their internal representations at the probing position during model inference.
% containing the last position in the prediction layer for next item predictions and all the preceding positions.
% All these residual stream activation values construct the training set of SAE.
With the same approach, we construct the valid and test set for RecSAE.

\subsection{Recommendation Models to be interpreted}
\label{subsection:models}

The RecSAE framework is capable of interpreting various kinds of recommendation models. 
For the following experiments, 
we interpret three types of recommendation models: general, graph-based, and sequential.

\begin{itemize}
    \item BPRMF~\cite{rendle2012bpr}: Byesian personalized ranking.
    \item LightGCN~\cite{he2020lightgcn}: graph convolution network for rec.
    \item SASRec~\cite{kang2018self}: self-attentive sequential recommendation.
    \item TiMiRec~\cite{wang2022target}: target-interest distillation framework for multi-interest sequential recommendation.
\end{itemize}

\subsection{Baseline}
\label{subsection:baseline}

RecSAE is a model-level interpretation framework that is not restricted to specific architectures. 
To the best of our knowledge, no existing work is applicable to interpreting general, graph-based, and sequential recommendation models in a unified manner.
% As shown in Table~\ref{table:related_work}, the only comparable generalized model-level baseline is ECF~\cite{ClusteringDu2023}, a clustering-based method that extracts taste clusters from interactions and item tags. 
% To further strengthen our comparison, we include the distillation method DistillEmb~\cite{DistillationZhang2020} as an additional baseline, though it is limited to jointly trained models. 
Therefore, for comparison purposes,  we probe the internal representations of base models for interpretation comparisons in Table~\ref{table:similarity_summary} and \ref{table:interpretation}. These baselines allow us to evaluate RecSAE’s effectiveness in performance and interpretability across diverse methods.

\subsection{Evaluation Metrics}
\label{subsection:metrics}
% \todo{to be modified}
RecSAE aims to interpret recommendation models, requiring that its latent reflect the internal states of the interpreted model. Thus, we evaluate both the reconstruction and interpretation performance of RecSAE latents.

For reconstruction, we evaluate the downstream recommendation performance by hit rate and NDCG at positions of 10 and 20 in test-all settings when the original representations are replaced with the reconstructions in the interpreted model, though this replacement is not needed during interpretation. The reconstruction results are shown in Section~\ref{section:reconstruction}.

For interpretation, we refer to related work~\cite{ClusteringDu2023} that measures the latent utilization through Silhouettes~\cite{rousseeuw1987silhouettes}~(a measure of how similar an object is to its own cluster) and the number of interpreted concepts within different confidence score ranges. 
Specifically, Silhouettes evaluates intra- and inter-concept similarity. The similarity is measured using the cosine distance between item embeddings in the interpreted models. Intra-concept similarity quantifies the average distances between the two most activated items within each concept. In addition, inter-concept similarity measures the average distances between pairs of distinct concepts. We calculate the above measures in concepts with a confidence score of 0.8 or higher.
High intra- and lower inter-concept similarity indicate that the interpreted concepts are well-defined and distinct. The interpretation results are shown in Section~\ref{section:interpretation}.

\subsection{Hyper-parameter}
\label{subsection:hyperparameter}
For recommendation models, we use the tuned settings in the ReChorus framework. The embedding size for recommenders is 64, so sparsity $k$ in RecSAE should be less than 64. We perform hyperparameter tuning over the scale size \{8, 16, 32\}, the sparsity k \{8, 16\}, learning rate \{1e-5, 5e-5, 1e-4\} and batch size \{4, 8, 16\}. The impact of scale size and sparsity k is presented in Section~\ref{section:recstruct_interpet}. For final experiments, we set the scale size to 16~(leading to 1024 latents) and k to 8 for reconstruction and interpretation balance. The detailed hyper-parameter settings for interpreting each model are reported in run.sh within the codebase.

In the following experiment sections, we present the interpretation performance, reconstruction performance, parameter sensitivity experiments, and targeted tuning.

\section{Interpretation Performance}
\label{section:interpretation}

RecSAE consists of two phases: the unsupervised learning phase, which forms the interpretable latents, and the interpretation phase, which generates semantic concepts based on the items that activate each latent. Thus, we evaluate both the quality of the learned latents and the interpreted concepts.

In the following sections, we:
1) Assess the quality of the latents learned by RecSAE;
2) Present the interpreted concepts, focusing on their quantity, activation patterns, concrete examples, and concept survey;
3) Conduct a human evaluation to assess the validity and clarity of the interpreted concepts produced by RecSAE.

% As RecSAE has two phases, the unsupervised learning phase to form the interpretable latents, and the interpretation phase to form the semantic concepts from items that activated each latents, we first evaluate the latent quality and then examine the interpreted concepts.

% In the following sections,
% we examine the quality of RecSAE learned latents in Section~\ref{subsection:latents};
% present the interpreted concepts with quantity and activation situations and concrete concepts showcases.
% Then, we conduct human evaluation to examine the interpreted concepts produced by RecSAE.

% In the following sections, we aim to answer these questions:
% Does the items activated the same latents share similarities?
% Are RecSAE learned latents 

\subsection{RecSAE Extracted Latents}
\label{subsection:latents}

\begin{table}[htbp]
\centering
\caption{Evaluation of RecSAE Latents: Intra- and Inter-Latent Similarity for RecSAE across Interpreted Models and Datasets. The best performance in each setting is highlighted in \textbf{bold}. Higher intra-latent similarity and lower inter-latent similarity show that RecSAE Learns valid and distinct latents.}
\label{table:similarity_summary}
 \resizebox{\linewidth}{!}{
\begin{tabular}{ll|cc|cc|cc|cc}
\toprule
\multirow{2}{*}{\textbf{\makecell[l]{Rec\\Model}}} & \multirow{2}{*}{\textbf{Method}} &\multicolumn{2}{c|}{\textbf{Amazon}} & \multicolumn{2}{c|}{\textbf{MovieLens}} & \multicolumn{2}{c|}{\textbf{LastFM}} & \multicolumn{2}{c}{\textcolor{black}{\textbf{Yelp}}}\\
          &  & Intra ↑ & Inter ↓ & Intra ↑ & Inter ↓ & Intra ↑ & Inter ↓ & \textcolor{black}{Intra ↑} & \textcolor{black}{Inter ↓} \\
\midrule
    & Base & 0.1764 & 0.2363 & 0.1804 & 0.4740 & 0.1437 & 0.1495 & \textcolor{black}{\textbf{0.5161}} &  	\textcolor{black}{0.6492}   \\
BPRMF     & RecSAE & \textbf{0.3056} & \textbf{0.1663} & \textbf{0.3795} & \textbf{0.2998} & \textbf{0.3058} & \textbf{0.1431}  & \textcolor{black}{0.3594} &	\textcolor{black}{\textbf{0.2626}}    \\
     & \textit{Improve} & \textit{+73.30\%} &	\textit{-29.65\%} &	\textit{+110.33\%} &	\textit{-36.75\%} &	\textit{+112.85\%} &	\textit{-4.24\%} & \textcolor{black}{\textit{-30.37\%}} & 	\textcolor{black}{\textit{-59.55\%}} \\
     \hline
  & Base  & 0.2161 & 0.2463 & 0.4430 & 0.2938 & 0.1818 & 0.1565  &  \textcolor{black}{0.3433 } & 	 \textcolor{black}{0.3235}  \\
LightGCN  & RecSAE & \textbf{0.4540} & \textbf{0.1496} & \textbf{0.6225} & \textbf{0.2704} & \textbf{0.4672} & \textbf{0.1365} &  \textcolor{black}{\textbf{0.6389}} & 	 \textcolor{black}{\textbf{0.2248}}    \\
     &  \textit{Improve} & \textit{+110.03\%}	& \textit{-39.23\%}	& \textit{+40.54\%}	& \textit{-7.97\%}	& \textit{+156.95\%} &	\textit{-12.79\%} &  \textcolor{black}{\textit{+86.08\%}} & 	 \textcolor{black}{\textit{-30.50\%}}  \\ \hline
    & Base & 0.3777 & 0.1808 & 0.4452 & 0.2071 & 0.2652 & 0.1755 & \textcolor{black}{0.1853}& 	\textcolor{black}{\textbf{0.1263}}    \\ 
SASRec    & RecSAE  & \textbf{0.7147} & \textbf{0.1587} & \textbf{0.6980} & \textbf{0.1745} & \textbf{0.5114} & \textbf{0.1579} & \textcolor{black}{\textbf{0.5836}} & 	\textcolor{black}{0.1307}   \\ 
       &  \textit{Improve} & \textit{+89.21\%}	& \textit{-12.20\%} &	\textit{+56.80\%} &	\textit{-15.76\%} &	\textit{+92.81\%} &	\textit{-10.03\%} &  \textcolor{black}{\textit{+215.00\%}} & 	 \textcolor{black}{\textit{+3.53\%}} \\ \hline
   & Base & 0.3418 & 0.1405 & 0.5119 & 0.2303 & 0.2332 & 0.1626  & \textcolor{black}{0.3282} &  	\textcolor{black}{\textbf{0.1261}}   \\
TiMiRec   & RecSAE  & \textbf{0.6288} & \textbf{0.1366} & \textbf{0.6564} & \textbf{0.1895} & \textbf{0.5332} & \textbf{0.1470} &  \textcolor{black}{\textbf{0.6635}}  	& \textcolor{black}{0.1375}   \\
      &  \textit{Improve} & \textit{+83.96\%}	& \textit{-2.77\%} &	\textit{+28.22\%} &	\textit{-17.71\%} &	\textit{+128.60\%} &	\textit{-9.56\%} &  \textcolor{black}{\textit{+102.18\%}} & 	 \textcolor{black}{\textit{+9.00\%}} \\
\bottomrule
\end{tabular}
}
\end{table}

We begin our evaluation of RecSAE by analyzing the quality of extracted latents after the unsupervised learning phase. Specifically, we aim to assess two essential properties of the learned latents:
(1) Intra-latent coherence, indicating that items highly activated by the same latent are similar to one another; and
(2) Inter-latent distinctiveness, meaning that items activated by different latents are meaningfully dissimilar.
To quantify these properties, we compute intra-latent similarity and inter-latent similarity metrics across different datasets and recommendation models. Higher intra-latent similarity suggests tighter clustering of semantically similar items under a shared latent, while lower inter-latent similarity implies that different latents capture non-overlapping aspects of item semantics.

Table~\ref{table:similarity_summary} reports the results for four representative recommendation models (BPRMF, LightGCN, SASRec, and TiMiRec) across \textcolor{black}{four} benchmark datasets (Amazon Grocery, MovieLens, LastFM and Yelp). For each model and dataset, we compare the base representation (i.e., the original model’s internal encoding) with the RecSAE learned sparse latents.

RecSAE significantly improves intra-latent similarity, confirming that items grouped under the same latent activation are more semantically cohesive. For instance, in the Amazon dataset, the intra-latent similarity of BPRMF increases from 0.1764 to 0.3056 (a +73.30\% improvement). 
Simultaneously, RecSAE reduces inter-latent similarity, ensuring greater separation between different latents. In the same setting, the inter-latent similarity for BPRMF drops from 0.2363 to 0.1663 (-29.65\%). This pattern of improved intra-latent cohesion and reduced inter-latent overlap holds consistently across all evaluated settings, including both general (e.g., BPRMF), graph-based (LightGCN), and sequential (e.g., SASRec, TiMiRec) models. 
There is one special case on BPRMF trained on Yelp, where both intra- and inter-latent similarities are generally lower.
Note that these similarity metrics represent only one aspect of evaluating RecSAE; the primary utility of RecSAE lies in producing interpretable concepts and enabling subsequent targeted tuning.
% Particularly notable is the performance on LastFM, where TiMiRec’s intra-latent similarity improves by +128.60\%, demonstrating RecSAE’s capacity to form semantically rich and well-partitioned latent spaces across different recommendation models and scenarios.

These results highlight RecSAE’s effectiveness in structuring the interpretable latent space purely from the internal representations of recommendation models through unsupervised learning. By consolidating semantically related items and disentangling unrelated patterns, RecSAE achieves two essential properties that underpin mono-semantic latents in recommendation systems.

% We first evaluate the quality of RecSAE latents after the unsupervised training process.
% The goal is to assess whether items activated by the same latent exhibit similarity and items associated with different latents are distinct.
% To this end, we use intra-latent and inter-latent similarity metrics to assess RecSAE's ability to identify valid and well-separated latents.
% As shown in Table~\ref{table:similarity_summary}, intra-latent similarity scores consistently exceed those of the base models, showing intra-latent activation items are similar. At the same time, inter-latent item similarity remains low compared to the base model, showing items activated at different latents are different. These results demonstrate that RecSAE learns valid and district latents across the four recommendation models and three datasets.

% The intra-latent and inter-latent similarity metrics examine RecSAE's ability to identify valid and distinct latents.
% As shown in Table~\ref{table:similarity_summary}, intra-concept similarities consistently surpass those of the base models. RecSAE's inter-concept similarity remains consistently low, indicating well-separations. 
% While some inter-concept similarities exceed those of the original model, this may be due to base models' limited set of concepts, with fewer than 10 confidence scores exceeding 0.8. Given the small number of concepts, a low inter-concept similarity is expected.

\textit{\textbf{Finding 1: RecSAE learns valid and distinct latents.}}

\subsection{RecSAE Interpreted Concepts}
\label{subsection:interpreted_concepts}

% After extracting the interpretable latents from the original model representations, we further transfer these latents into semantic concepts for interpretation. In the following subsections, we present a quantitative overview of interpretation results, a qualitative case study of interpreted concepts and a concept survey that shows the overall pattern and coverage

After extracting interpretable latents from the internal representations of recommendation models, we further transform these latents into human-readable semantic concepts. In this section, we present analysis of this interpretation process, including quantitative overview of interpretation performance, latent activation situations, case study of representative concepts, and broader concept surveys that highlight the effectiveness, coverage, and emerging patterns of interpretation.

\subsubsection{Overall Performance}
\label{subsection:overall_performance}

\begin{table}[htbp]
\centering
\caption{RecSAE Interpretation Situation. We use the TF-IDF algorithm and a large language model, Llama-3-8B-Instruct, to map the latents to corresponding concepts. We then report both the number of \textcolor{black}{unique} verified concepts and the total number of \textcolor{black}{unique} concepts. For concept verification, we adopt "P$\geq$0.8", where "P" represents $Precision_{concept}$, as defined in Equation \ref{equation:precision}. The results show that RecSAE with the LLM interpretation produces a greater number of concepts.}
\label{table:interpretation}
 \resizebox{\linewidth}{!}{
\begin{tabular}{ll|rrrr|rrrr|rrrr|rrrr}
\toprule
\multirow{3}{*}{\textbf{\makecell[l]{Rec\\Model}}} & \multirow{3}{*}{\textbf{\makecell{Method}}} &\multicolumn{4}{c|}{\textbf{Amazon}} & \multicolumn{4}{c|}{\textbf{MovieLens}} & \multicolumn{4}{c|}{\textbf{LastFM}} & \multicolumn{4}{c}{\textbf{\textcolor{black}{Yelp}}} \\
% & &  \multicolumn{2}{c}{Base} & \multicolumn{2}{c|}{RecSAE} &  \multicolumn{2}{c}{Base}& \multicolumn{2}{c|}{RecSAE} &  \multicolumn{2}{c}{Base} & \multicolumn{2}{c}{RecSAE}  \\
& &  \multicolumn{2}{c}{\textcolor{black}{Base}} & \multicolumn{2}{c|}{\textcolor{black}{RecSAE}} &  \multicolumn{2}{c}{\textcolor{black}{Base}}& \multicolumn{2}{c|}{\textcolor{black}{RecSAE}} &  \multicolumn{2}{c}{\textcolor{black}{Base}} & \multicolumn{2}{c|}{\textcolor{black}{RecSAE}} &  \multicolumn{2}{c}{\textcolor{black}{Base}} & \multicolumn{2}{c}{\textcolor{black}{RecSAE}}    \\

           &  & P$\geq$0.8 & All & P$\geq$0.8 & All & P$\geq$0.8 & All  & P$\geq$0.8 & All & P$\geq$0.8 & All  & P$\geq$0.8 & All & P$\geq$0.8 & All  & P$\geq$0.8 & All \\
\midrule

      & \textcolor{black}{TF-IDF} &  \textcolor{black}{1}    & \textcolor{black}{56}   & \textcolor{black}{28}   & \textcolor{black}{466}  & \textcolor{black}{0}    & \textcolor{black}{58}   & \textcolor{black}{14}   & \textcolor{black}{505}  & \textcolor{black}{0}    & \textcolor{black}{53}   & \textcolor{black}{27}   & \textcolor{black}{378} & \textcolor{black}{\textbf{9}} &	\textcolor{black}{22} &	\textcolor{black}{28} &	\textcolor{black}{59}  \\
BPRMF     & \textcolor{black}{LLM} &   \textcolor{black}{\textbf{5}}    & \textcolor{black}{\textbf{64}}   & \textcolor{black}{\textbf{110}}  & \textcolor{black}{\textbf{1009}} & \textcolor{black}{\textbf{2}}     & \textcolor{black}{\textbf{64}}   & \textcolor{black}{\textbf{138}}  & \textcolor{black}{\textbf{1019}} & \textcolor{black}{\textbf{4}}    & \textcolor{black}{\textbf{64}}   & \textcolor{black}{\textbf{138}}  & \textcolor{black}{\textbf{1019}} & \textcolor{black}{5} &	\textcolor{black}{\textbf{64}} &	\textcolor{black}{\textbf{63}} &	\textcolor{black}{\textbf{672}} \\

      &  \textit{Improve} & \textcolor{black}{\textit{4.00}} & \textcolor{black}{\textit{0.14}} & \textcolor{black}{\textit{2.93}} & \textcolor{black}{\textit{1.17}} & \textcolor{black}{\textit{-}}     & \textcolor{black}{\textit{0.10}} & \textcolor{black}{\textit{8.86}} & \textcolor{black}{\textit{1.02}} & \textcolor{black}{\textit{-}}    & \textcolor{black}{\textit{0.21}} & \textcolor{black}{\textit{4.11}} & \textcolor{black}{\textit{1.70}} & \textcolor{black}{\textit{-0.44}} & 	\textcolor{black}{\textit{1.91}} & 	\textcolor{black}{\textit{1.25}} & 	\textcolor{black}{\textit{10.39}}  \\
     \hline

        & \textcolor{black}{TF-IDF}  &  \textcolor{black}{0}             & \textcolor{black}{59}            & \textcolor{black}{49}            & \textcolor{black}{469}           & \textcolor{black}{7}              & \textcolor{black}{46}            & \textcolor{black}{42}            & \textcolor{black}{409}           & \textcolor{black}{0}             & \textcolor{black}{54}            & \textcolor{black}{67}            & \textcolor{black}{439}    &   \textcolor{black}{\textbf{9}}    &   \textcolor{black}{22}    &   \textcolor{black}{\textbf{83}}    &   \textcolor{black}{222}     \\
LightGCN  & \textcolor{black}{LLM} &   \textcolor{black}{\textbf{2}}    & \textcolor{black}{\textbf{64}}   & \textcolor{black}{\textbf{130}}  & \textcolor{black}{\textbf{1020}} & \textcolor{black}{\textbf{8}}     & \textcolor{black}{\textbf{64}}   & \textcolor{black}{\textbf{92}}   & \textcolor{black}{\textbf{701}}  & \textcolor{black}{\textbf{3}}    & \textcolor{black}{\textbf{64}}   & \textcolor{black}{\textbf{138}}  & \textcolor{black}{\textbf{1024}}    &   \textcolor{black}{2}    &   \textcolor{black}{\textbf{64}}    &   \textcolor{black}{74}    &   \textcolor{black}{\textbf{831}} \\
      &  \textit{Improve} & \textcolor{black}{\textit{-}}    & \textcolor{black}{\textit{0.08}} & \textcolor{black}{\textit{1.65}} & \textcolor{black}{\textit{1.17}} & \textcolor{black}{\textit{0.14}}  & \textcolor{black}{\textit{0.39}} & \textcolor{black}{\textit{1.19}} & \textcolor{black}{\textit{0.71}} & \textcolor{black}{\textit{-}}    & \textcolor{black}{\textit{0.19}} & \textcolor{black}{\textit{1.06}} & \textcolor{black}{\textit{1.33}}    &   \textcolor{black}{\textit{-0.78}}    &   \textcolor{black}{\textit{1.91}}    &   \textcolor{black}{\textit{-0.11}}    &   \textcolor{black}{\textit{2.74}}  \\ \hline

           & \textcolor{black}{TF-IDF} &  \textcolor{black}{7}             & \textcolor{black}{53}            & \textcolor{black}{131}           & \textcolor{black}{468}           & \textcolor{black}{\textbf{7}}     & \textcolor{black}{50}            & \textcolor{black}{53}            & \textcolor{black}{508}           & \textcolor{black}{2}             & \textcolor{black}{53}            & \textcolor{black}{78}            & \textcolor{black}{506}    &   \textcolor{black}{\textbf{4}}    &   \textcolor{black}{20}    &   \textcolor{black}{37}    &   \textcolor{black}{65}          \\

SASRec & \textcolor{black}{LLM} & \textcolor{black}{\textbf{10}}   & \textcolor{black}{\textbf{64}}   & \textcolor{black}{\textbf{157}}  & \textcolor{black}{\textbf{1004}} & \textcolor{black}{4}              & \textcolor{black}{\textbf{64}}   & \textcolor{black}{\textbf{111}}  & \textcolor{black}{\textbf{970}}  & \textcolor{black}{\textbf{8}}    & \textcolor{black}{\textbf{64}}   & \textcolor{black}{\textbf{111}}  & \textcolor{black}{\textbf{970}}            & \textcolor{black}{1}            & \textcolor{black}{\textbf{64}}            & \textcolor{black}{\textbf{173}}            & \textcolor{black}{\textbf{1017}} \\

       & \textit{Improve} &   \textcolor{black}{\textit{0.43}} & \textcolor{black}{\textit{0.21}} & \textcolor{black}{\textit{0.20}} & \textcolor{black}{\textit{1.15}} & \textcolor{black}{\textit{-0.43}} & \textcolor{black}{\textit{0.28}} & \textcolor{black}{\textit{1.09}} & \textcolor{black}{\textit{0.91}} & \textcolor{black}{\textit{3.00}} & \textcolor{black}{\textit{0.21}} & \textcolor{black}{\textit{0.42}} & \textcolor{black}{\textit{0.92}}   & \textcolor{black}{\textit{-0.75}}   & \textcolor{black}{\textit{2.20}}   & \textcolor{black}{\textit{3.68}}   & \textcolor{black}{\textit{14.65}}  \\  \hline

         & \textcolor{black}{TF-IDF} & \textcolor{black}{2}             & \textcolor{black}{55}            & \textcolor{black}{67}            & \textcolor{black}{482}           & \textcolor{black}{4}              & \textcolor{black}{45}            & \textcolor{black}{34}            & \textcolor{black}{485}           & \textcolor{black}{0}             & \textcolor{black}{56}            & \textcolor{black}{76}            & \textcolor{black}{506}   & \textcolor{black}{\textbf{5}}   & \textcolor{black}{52}   & \textcolor{black}{67}   & \textcolor{black}{324}          \\

TiMiRec & \textcolor{black}{LLM} &  \textcolor{black}{\textbf{10}}   & \textcolor{black}{\textbf{64}}   & \textcolor{black}{\textbf{193}}  & \textcolor{black}{\textbf{994}}  & \textcolor{black}{\textbf{7}}     & \textcolor{black}{\textbf{64}}   & \textcolor{black}{\textbf{81}}   & \textcolor{black}{\textbf{804}}  & \textcolor{black}{\textbf{11}}   & \textcolor{black}{\textbf{64}}   & \textcolor{black}{\textbf{227}}  & \textcolor{black}{\textbf{997}}   & \textcolor{black}{3}   & \textcolor{black}{\textbf{64}}   & \textcolor{black}{\textbf{141}}   & \textcolor{black}{\textbf{956}}  \\

       & \textit{Improve}  & \textcolor{black}{\textit{4.00}} & \textcolor{black}{\textit{0.16}} & \textcolor{black}{\textit{1.88}} & \textcolor{black}{\textit{1.06}} & \textcolor{black}{\textit{0.75}}  & \textcolor{black}{\textit{0.42}} & \textcolor{black}{\textit{1.38}} & \textcolor{black}{\textit{0.66}} & \textcolor{black}{\textit{-}}    & \textcolor{black}{\textit{0.14}} & \textcolor{black}{\textit{1.99}} & \textcolor{black}{\textit{0.97}}   & \textcolor{black}{\textit{-0.40}}   & \textcolor{black}{\textit{0.23}}   & \textcolor{black}{\textit{1.10}}   & \textcolor{black}{\textit{1.95}}
 \\

\bottomrule
\end{tabular}
}
\end{table}

Table~\ref{table:interpretation} presents the quantitative evaluation results of the interpreted concepts. We compare the representations from the original recommendation models with the latents extracted by RecSAE, applying two independent interpretation pipelines: a TF-IDF–based term-level statistical method and a large language model (Llama-3-8B-Instruct~\cite{llama3modelcard}) capable of semantic summarizations.

\textcolor{black}{Across all datasets and base recommendation models, LLMs generally yield a larger number of verified semantic concepts than the TF-IDF method. This superior performance reflects their stronger semantic understanding and ability to extract coherent concepts beyond surface-level term-frequency signals.
In addition, the improvements of RecSAE over the base recommendation models are evident under both interpretation methods, demonstrating that RecSAE latents possess rich interpretability, even when evaluated using a purely term-level statistical approach.}

Furthermore, sequential recommendation models (SASRec and TiMiRec) generally produce more interpretable and verifiable latents than non-sequential models (BPRMF and LightGCN). These interpretation results align with their superior recommendation accuracy performance.
The interpretation results from RecSAE provide additional evidence that these models are more effective at capturing meaningful patterns from collaborative signals in purely ID-based user-item interaction logs.

% Table~\ref{table:interpretation} reports quantitative evaluations of the interpreted concepts. We evaluate both interpretations of the original model representations and the latents extracted by RecSAE, using two independent interpretation pipelines: a TF-IDF–based method and an open-source, locally deployed large language model (Llama-3-8B-Instruct~\cite{llama3modelcard}).

% Across all three datasets, RecSAE consistently produces a greater number of verified concepts compared to the four original models. 
% When comparing interpretation methods, the LLM-based approach generally yields more verified concepts than the TF-IDF method, reflecting its stronger capacity for semantic understanding beyond surface-level term frequency algorithms.
% Among the recommendation models, sequential recommenders (SASRec and TiMiRec) yield more interpretable and verifiable latents than non-sequential methods (BPRMF and LightGCN), highlighting their superior capacity for capturing meaningful patterns from ID-based user-item interaction data.

\textit{\textbf{Finding 2: RecSAE latents can be interpreted into verified concepts with both TF-IDF and LLM-based methods.}}

% Nevertheless, even the strongest baseline exhibits limited interpretability compared with RecSAE, whose latents support fine-grained, semantically coherent interpretations across both datasets and both interpretation methods.

% Table~\ref{table:interpretation} presents the quantitative measurements of the overall interpreted concept situations. We use both TF-IDF and a open-sourced local-deployed large language model (Llama-3-8B-Instruct) to interpret the original model representations and RecSAE extracted latents.
% RecSAE latents can be interpreted into more valid concepts than various base models. 
% Compared to non-sequential models (BPRMF, LightGCN), sequential models (SASRec, TiMiRec) are interpreted into more verified latents across both datasets and interpretation methods, showing their superior encoding performance.
% While the four recommendation models exhibit limited interpretability in their representations, RecSAE extracts latents that offer fine-grained interpretations with high confidence.

\subsubsection{\textcolor{black}{Confidence Score Distribution}}
\label{subsubsection:confidence_score}

\begin{figure}[htbp]
    \centering
     \includegraphics[width=0.32\linewidth]{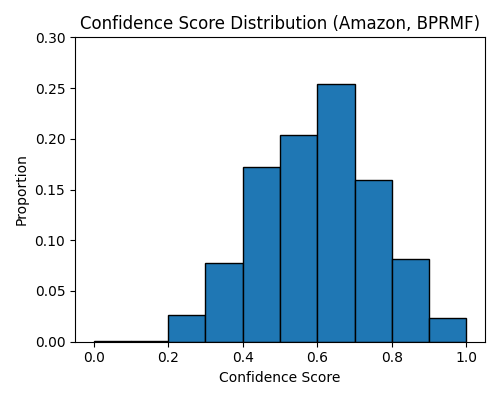}
      \includegraphics[width=0.32\linewidth]{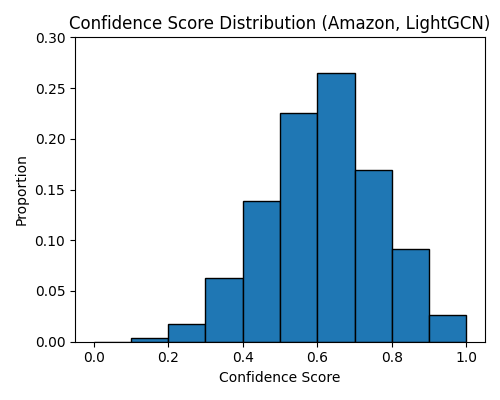}
      \includegraphics[width=0.32\linewidth]{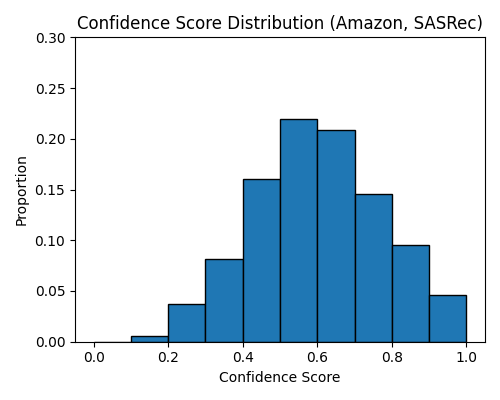}
    \caption{\textcolor{black}{Confidence Score Distribution across Interpreted Latents on BPRMF, LightGCN and SASRec on Amazon dataset.}}
    \label{figure:confidence_score_distribution}
\end{figure}

\textcolor{black}{The above results summarize the overall performance across four models and four datasets. To gain deeper insights into the \textit{confidence score} rather than only snapshots at thresholds such as 0.8 or the entire range, we plot the distributions of latent confidence scores in Figure~\ref{figure:confidence_score_distribution}. Specifically, we visualize the distributions on the Amazon dataset across three recommendation models. The results show that the majority of latents exhibit confidence scores above 0.6. Moreover, as the base recommendation model becomes stronger, the proportion of latents with confidence scores exceeding 0.9 increases. These observations indicate that the learned latents are interpretable with substantial confidence, and that their qualities are closely aligned with the modeling capability of the underlying base recommenders.}

\subsubsection{Case Study}
\label{subsection:case_study}

We provide examples concepts inferred by RecSAE to demonstrate both the inherent interpretability of its latents and the validity of the interpreted concepts.
The full latent information and corresponding concepts of RecSAE is presented in our codebase\footnote{https://github.com/Alice1998/RecSAE/tree/main/results}. 
% \todo{to check!}

% We provide examples of RecSAE concepts to show the inherent interpretability in the RecSAE latents and the reasonable of interpreted concepts.

The interpretation cases for SASRec on \textcolor{black}{four} datasets are presented in Table~\ref{table:examples}.
We present the latent information (latent id, corresponding activated item ids and titles) and the interpreted concepts (with TF-IDF and LLM methods for interpretation).
Notably, items that are activated by the same latent exhibit strong semantic similarities. In the Amazon grocery dataset, the concepts range from specific food types to particular ingredients. In MovieLens, the latents span from specific movie series to various genres. In LastFM, the interpreted concepts include different singers, albums and music types. These cases collectively demonstrate that RecSAE effectively captures semantically meaningful concepts within the internal representations of recommendation models.

Importantly, these item groups that activated the same latent emerge solely from the internal representations learned by the recommendation models and are extracted by RecSAE through unsupervised training. The above process does not incorporate any semantic information. It is particularly noteworthy that such interpretable concepts arise within ID-based recommendation models trained purely on collaborative user–item interaction data. This highlights the capacity of collaborative filtering methods to uncover structured semantic patterns from interaction signals alone.
RecSAE, through its unsupervised framework, sheds light on these latent concepts, revealing the rich, human-understandable structures embedded within otherwise opaque recommendation models.

% Note that these item clusters are formed by the internal representation of the recommendation models, and extracted by RecSAE through unsupervised learning. The entire process does not utilize any semantic information.
% It is interesting to see these meaningful concepts are encoded in the ID-based recommendation model learned with collaborative signals in user item interaction logs.
% This underscores the remarkable ability of collaborative filtering to capture meaningful information from data.
% RecSAE, employing an unsupervised training approach, reveals these concepts embedded in previously opaque models.

% While some results may seem to reflect item clusters based on their descriptions, these "clusters" are actually formed by the interpreted model.
% These purely ID-based models encode human-understandable concepts within their activations after training exclusively on user-item interactions, without any semantic information about items or users. 
% This underscores the remarkable ability of collaborative filtering to capture meaningful information from data.
% RecSAE, employing an unsupervised training approach, reveals these concepts embedded in previously opaque models.

% Table~\ref{table:examples} shows the randomly selected examples of Latent information and corresponding Concepts.
% Due to space limitations, we show the randomly selected six cases across datasets. The full interpretation information is recorded in our codebase\footnote{\todo{log link}}.

\begin{table}[htbp]
\caption{Examples of RecSAE Latent information and Corresponding Concepts derived from SASRec model on \textcolor{black}{four} datasets. The pure ID-based model, trained without semantic information, encodes human-understandable concepts within its representations. \textcolor{black}{The TF-IDF concepts are the Top-3 terms.} The full logs of all latents are presented in our codebase.}
    \label{table:examples}
    \centering
    \resizebox{\textwidth}{!}{
    \begin{tabular}{lllll}
    % \begin{tabular}{lcc>{\makecell[l]{}}ccl}
    \toprule
 \makecell[l]{Dataset\\-LID} & \makecell{IID} & Item Title &\makecell[l]{TF-IDF\\Concept} & LLM Concept \\ \hline

% \multirow{6}{*}{Amazon} & 
\multirow{3}{*}{\makecell{A-193}} & 4582 &	Betty Crocker \textbf{Gluten Free Brownie Mix},...	 &  \multirow{3}{*}{\makecell[l]{\textbf{Gluten},\\\textbf{Free},\\Mix}} & \multirow{3}{*}{\makecell[l]{\textbf{gluten-free baked goods},\\with a particular focus on\\mixes for baked goods}} \\
% such as\\brownies, cakes, cookies, and pancake}}\\
 & 4575	& Betty Crocker \textbf{Gluten Free} Chocolate Chip \textbf{Cookie Mix}, ... &	& \\
 & 5882 &	Bisquick \textbf{Pancake and Baking Mix}, \textbf{Gluten-Free}, ... &	 & \\ 
 \cmidrule(lr){2-5}

\multirow{3}{*}{\makecell{A-271}} &	7644 &	Happy \textbf{Baby} Happy Munchies Rice Cakes Apple ...	 & \multirow{3}{*}{\makecell[l]{\textbf{Baby},\\Happy,\\Cake}} &  \multirow{3}{*}{\makecell[l]{\textbf{baby food or snacks},\\ specifically organic and\\health options for infants}}\\
% and toddlers
&	5060	& Happy Baby Organic \textbf{Baby Food} 3 Homestyle Meals,... &	 &  \\
 &	5057 &	Happy Tot Organic \textbf{Baby Food}, Sweet Potato,... &	& \\    \cmidrule(lr){2-5} 
 
 \multirow{3}{*}{\makecell{A-275}} & 7890  &	... \textbf{Decaffeinated} \textbf{Coffee} ... Keurig K-Cup Brewers
 &  \multirow{3}{*}{\makecell[l]{Keurig,\\\textbf{Coffee},\\Brewers}} & \multirow{3}{*}{\makecell[l]{\textbf{coffee}-based,\\particularly single-serve cups\\with a focus on \textbf{decaf}}} \\
% such as\\brownies, cakes, cookies, and pancake}}\\
 & 7968	& The Organic \textbf{Coffee} Co. \textbf{Decaf} Gorilla Coffee ...
 &	& \\
 & - &	- &	 & \\ 
 \hline

% \multirow{6}{*}{\makecell{Movie-\\Lens}} &
% \multirow{3}{*}{\makecell{M-29}} & 1085 &	\textbf{Star Trek} IV: The Voyage Home (Action, Adventure, Sci-Fi) &	\multirow{3}{*}{\makecell[l]{\textbf{Trek},\\\textbf{Star},\\Adventure}} & \multirow{3}{*}{\makecell[l]{part of a science fiction franchise, \\ particularly the \textbf{Star Trek franchise}, \\ or have a strong science fiction element}} \\
%  & 1069 &	\textbf{Star Trek}: First Contact (Action, Adventure, Sci-Fi) &	 & \\
%  & 1083 &	\textbf{Star Trek}: The Wrath of Khan (Action, Adventure, Sci-Fi)	&  & \\   \cmidrule(lr){2-5} 
% trek', 'star', 'adventure'

%   \multirow{3}{*}{\makecell{M-684}}
% & 511	& Aladdin (\textbf{Animation}, \textbf{Children's}, Comedy, Musical) &	\multirow{3}{*}{\makecell[l]{Bambi,\\\textbf{Animation},\\\textbf{Children}}}  & \multirow{3}{*}{\makecell[l]{feature \textbf{child-friendly} 
% , such as \\ animation, children's stories, or musicals,  \\ often with a strong emotional}}\\
% & 1551 &	Bambi (\textbf{Animation}, \textbf{Children's}) & 	 & \\
% & 1311	& Anastasia (\textbf{Animation}, \textbf{Children's}, Musical)	& &  \\   \hline
% 588
% Antz (1998) (Animation, Children's) 2294
% Robin Hood (1973) (Animation, Children's) 3034
% bambi', 'animation', 'children'

% trek', 'fi', 'sci', 'predator
% 1376 - Star Trek IV: The Voyage Home (1986) (Action, Adventure, Sci-Fi)
% 1356- Star Trek: First Contact (1996) (Action, Adventure, Sci-Fi)
% 1374-  Star Trek: The Wrath of Khan (1982) (Action, Adventure, Sci-Fi)

\multirow{3}{*}{\makecell{M-204}} & 2496 &	Blast from the Past (1999) (\textbf{Comedy}, \textbf{Romance}) &	\multirow{3}{*}{\makecell[l]{\textbf{Romance},\\Seattle,\\Sleepless}} & \multirow{3}{*}{\makecell[l]{\textbf{romantic comedies},\\featuring light-hearted,\\ humorous storylines}} \\
 & 2671 &	Notting Hill (1999) (\textbf{Comedy}, \textbf{Romance}) &	 & \\
 & 539 &	Sleepless in Seattle (1993) (\textbf{Comedy}, \textbf{Romance})	&  & \\   \cmidrule(lr){2-5}

\multirow{3}{*}{\makecell{M-612}} & 1376 &	\textbf{Star Trek} IV: The Voyage Home (Action, Adventure, Sci-Fi) &	\multirow{3}{*}{\makecell[l]{\textbf{Trek},\\\textbf{Sci-Fi},\\\textbf{Predator}}} & \multirow{3}{*}{\makecell[l]{the \textbf{Star Trek franchise}, \\or have a strong\\science fiction element}} \\
 & 1356 &	\textbf{Star Trek}: First Contact (Action, Adventure, Sci-Fi) &	 & \\
 & 1374 &	\textbf{Star Trek}: The Wrath of Khan (Action, Adventure, Sci-Fi)	&  & \\   \cmidrule(lr){2-5} 
 % part of a science fiction franchise,\\particularly the \textbf{Star Trek franchise}, \\or have a strong science fiction element

  \multirow{3}{*}{\makecell{M-684}}
& 588	& Aladdin (1992) (\textbf{Animation}, \textbf{Children's}, Comedy, Musical) &	\multirow{3}{*}{\makecell[l]{\textbf{Animation},\\Aladdin,\\\textbf{Children}}}  & \multirow{3}{*}{\makecell[l]{feature \textbf{child-friendly},\\ 
such as animation, children\\stories, or musicals}}\\
& 2294 &	Antz (1998) (\textbf{Animation}, \textbf{Children's}) & 	 & \\
& 3034	& Robin Hood (1973) (\textbf{Animation}, \textbf{Children's})	& &  \\   \hline

% \multirow{6}{*}{LastFM} & 

\multirow{3}{*}{\makecell{L-254}}
&  459  &	Adios, in album HUSH, by EVERGLOW (pop, \textbf{k-pop} ...)	& \multirow{3}{*}{\makecell[l]{\textbf{k-pop},\\merry,\\room}} & \multirow{3}{*}{\makecell[l]{high-energy, upbeat \textbf{K-pop}\\ music with strong electronic\\and danceable elements ...}}\\
& 26336	 &	Snapping, in album Flourishing, by CHUNG HA (... \textbf{k-pop}) & 	& \\ 
&	25774 &	FANCY, in album FANCY YOU, by TWICE (\textbf{k-pop})	&  & \\   \cmidrule(lr){2-5} 

\multirow{3}{*}{\makecell{L-260}}
&   21101 &	I'll Follow The Sun ..., by \textbf{The Beatles}	& \multirow{3}{*}{\makecell[l]{\textbf{Beatles},\\Remastered,\\Rock}} & \multirow{3}{*}{\makecell[l]{from the \textbf{Beatles}, \\with songs considered their\\most iconic and well-known}}\\
&	9564 &	Norwegian Wood (This Bird Has Flown) ..., by \textbf{The Beatles} & 	& \\ 
&	4795 &	Strawberry Fields Forever ..., ..., by \textbf{The Beatles} ...	&  & \\   \cmidrule(lr){2-5} 
% 'beatles', 'remastered', 'rock'

 \multirow{3}{*}{\makecell{L-804}}
&   26791 &	Traust,... Futha, by Heilung (... dark folk, \textbf{nordic folk}) &	\multirow{3}{*}{\makecell[l]{Yggdrasil,\\Runaljod,\\\textbf{Nordic}}} & \multirow{3}{*}{\makecell[l]{connect to \textbf{Nordic culture},\\particularly folk music with\\traditional instruments}}\\ &	9656 &	Fehu,... Runaljod–Yggdrasil, by Wardruna (folk, \textbf{nordic folk}) &	 & \\ 
&	9660 &	Solringen,... Runaljod–Yggdrasil, by Wardruna (\textbf{nordic folk})	&  & \\ \hline

\multirow{3}{*}{\makecell{Y-84}}
&  1372  &	Bartaco, in \textbf{Nashville}, TN, (\textbf{Mexican}, Nightlife, Bars)	& \multirow{3}{*}{\makecell[l]{Chicken,\\\textbf{Nashville},\\TN}} & \multirow{3}{*}{\makecell[l]{\textbf{Southern-style restaurants}\\and bars that serve traditional\\\textbf{American} comfort food}} \\
& 2674	 &	Hattie B's Hot Chicken, in \textbf{Nashville}, TN (\textbf{American}, \textbf{Southern}) & 	& \\ 
& 13966 &	The Stillery, in \textbf{Nashville}, TN (\textbf{American})	&  & \\   \cmidrule(lr){2-5}

\multirow{3}{*}{\makecell{Y-352}}
&  3778  &	\textbf{Café} Du Monde, in \textbf{New Orleans}, LA (\textbf{Cafes}, Restaurants)	& \multirow{3}{*}{\makecell[l]{Shops,\\\textbf{Orleans},\\LA}} & \multirow{3}{*}{\makecell[l]{\textbf{Coffee} shops\\that serve traditional\\\textbf{New Orleans}-style coffee...}} \\

& 4391	 &	Dong Phuong, in \textbf{New Orleans}, LA (Breakfast, Brunch) & 	& \\ 

& 11395 &	Bearcat \textbf{Cafe}, in \textbf{New Orleans}, LA (\textbf{Coffee}, Breakfast, Brunch)	&  & \\   \cmidrule(lr){2-5}

\multirow{3}{*}{\makecell{Y-846}}
&  10553  &	Talula's Garden, in \textbf{Philadelphia}, PA (Restaurants)	& \multirow{3}{*}{\makecell[l]{\textbf{Bars},\\Parc,\\\textbf{Philadelphia}}} & \multirow{3}{*}{\makecell[l]{\textbf{Fine dining restaurants} that\\serve French-inspired cuisine\\with a focus on \textbf{wine}...}} \\
& 10641	 &	Parc, in \textbf{Philadelphia}, PA (\textbf{French}, \textbf{Wine Bars}, \textbf{Nightlife}) & 	& \\ 
& 13447 &	Suraya, in \textbf{Philadelphia}, PA (\textbf{Beer Gardens}, \textbf{Nightlife})	&  & \\

\bottomrule
\end{tabular}
}
\end{table}

% \textbf{\textit{Finding 3: RecSAE interpreted concepts are human-readable and reasonable.}}
\textbf{\textit{Finding 3: RecSAE interprets human-readable concepts from ID-based recommendation models trained without any semantic information.}}

\begin{figure}[htbp]
    \centering
    \includegraphics[width=0.45\linewidth]{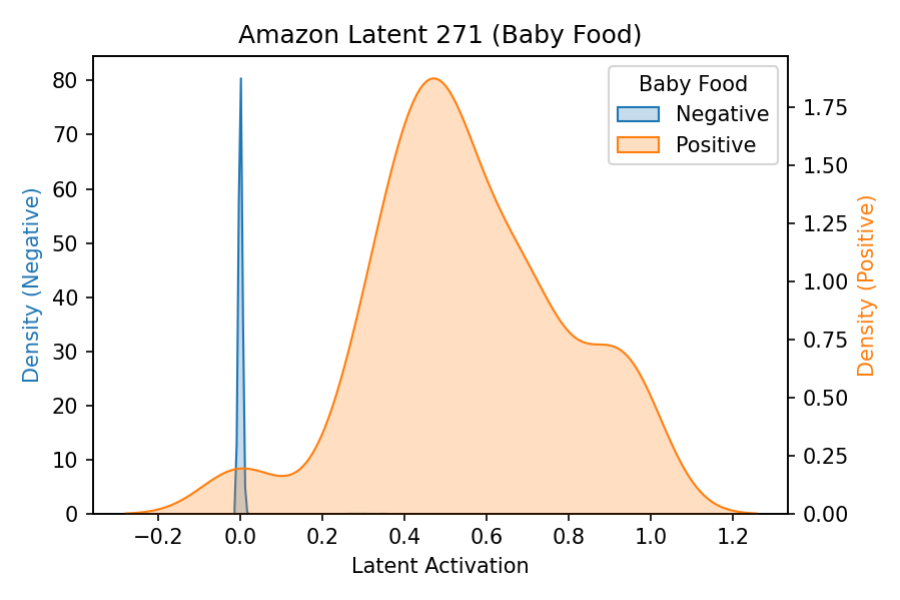}
    \includegraphics[width=0.45\linewidth]{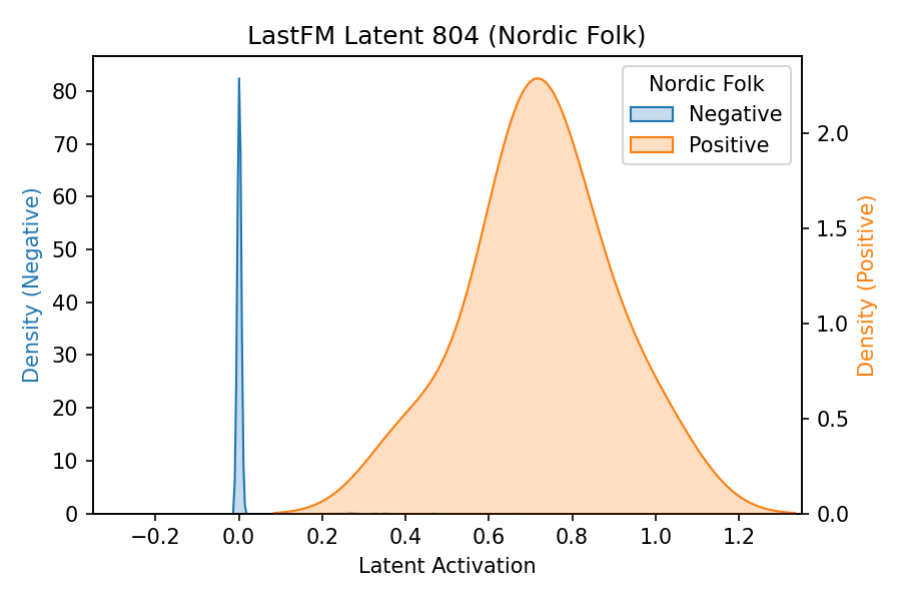}
    \caption{RecSAE Latent Activation Distribution of SASRec. Cases are categorized as negative (blue) or positive (orange) based on the presence of the corresponding concepts. Results show that positive cases exhibit higher activations in the corresponding latents.}
    \label{figures:activation_cases}
\end{figure}

We further present the randomly selected cases on latent activation distributions across datasets, as shown in Figure~\ref{figures:activation_cases}. The distributions are grouped based on whether the corresponding interpreted concept is present in the activated items. In positive cases, where the concept is present, the associated latent exhibits significantly higher activation values compared to negative cases, where the concept is absent. This clear separation indicates that RecSAE successfully aligns latent activations with the presence of corresponding concepts.

These results provide empirical evidence of both the necessity and sufficiency of latent activations for concept presence. Latents are strongly activated when their associated concepts appear (sufficiency), and are minimally activated otherwise (necessity). This further validates the interpretability and semantic consistency of RecSAE interpretations on recommendation models.

% We further demonstrate case study of latent activated distributions on the three datasets, grouped by whether the corresponding concepts are present in the activated items in Figure~\ref{figures:activation_cases}.
% Positive cases—those containing the corresponding concept—trigger higher activations in the associated latent compared to negative cases. This demonstrates that RecSAE successfully aligns latent activations with interpretable concept presence, validating the semantic consistency of its representations.

\textbf{\textit{Finding 4: The presence of concepts in recommendation model outputs corresponds to distinct patterns of associated latent activations.}}

% \textbf{\textit{Findings 3: Case study shown}}

% The above results show the case study of interpretation results. We then plot the overall situation of interpreted concepts.

\subsubsection{Concept Survey}
\label{subsection:concept_survey}

The preceding case study demonstrates that RecSAE is capable of extracting highly interpretable latent concepts from recommendation models. These concepts are human-readable, semantically coherent and easy to verify. To further assess the interpretability provided by RecSAE, we conduct a comprehensive survey of the interpreted concepts, illustrating their potential to reveal meaningful insights into the model's capabilities.

\begin{figure}[htbp]
    \centering
    
    % \includegraphics[width=0.45\linewidth]{figures/word_cloud_BPRMF.pdf}
    % \includegraphics[width=0.45\linewidth]{figures/word_cloud_SASRec.pdf}
    % \caption{Concept Survey on BPRMF~(left, NDCG@10=0.0200) and SASRec~(right, NDCG@10=0.0367) trained on Amazon. RecSAE interprets more diverse concepts in models with stronger recommendation performance.}
     \includegraphics[width=0.45\linewidth]{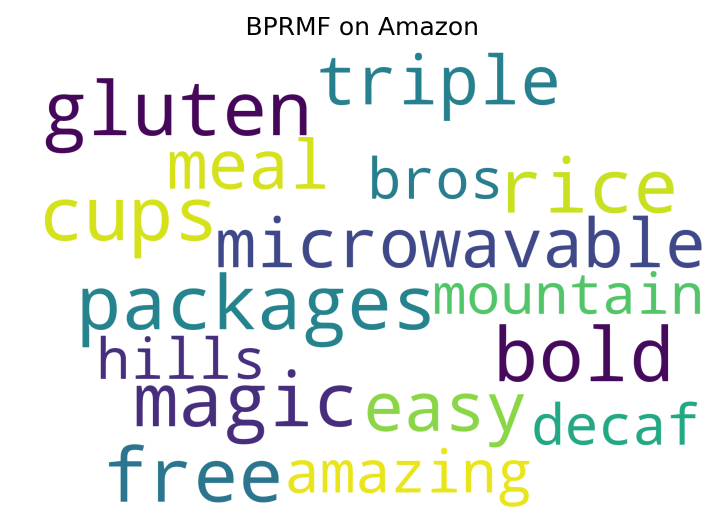}
    \includegraphics[width=0.45\linewidth]{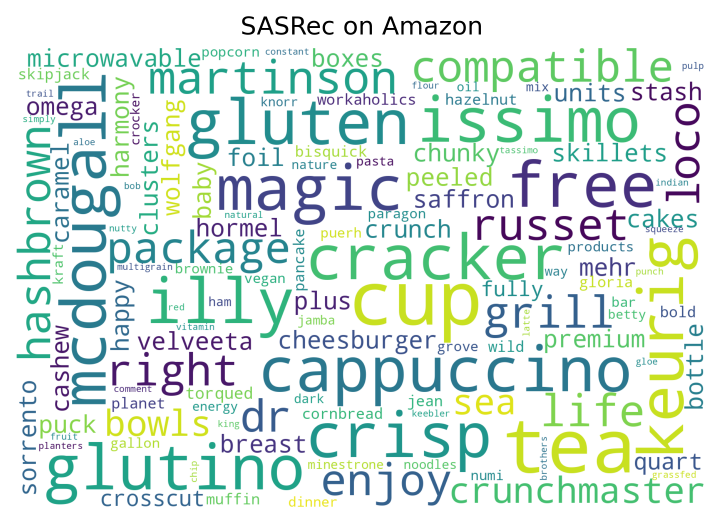}

    \includegraphics[width=0.45\linewidth]{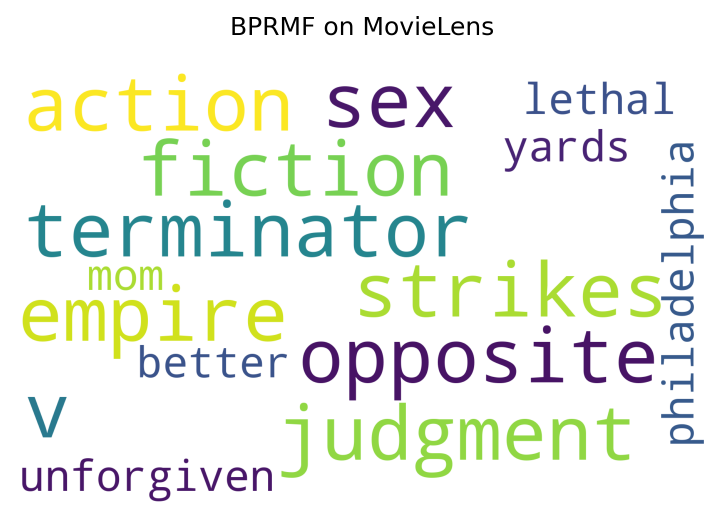}
    \includegraphics[width=0.45\linewidth]{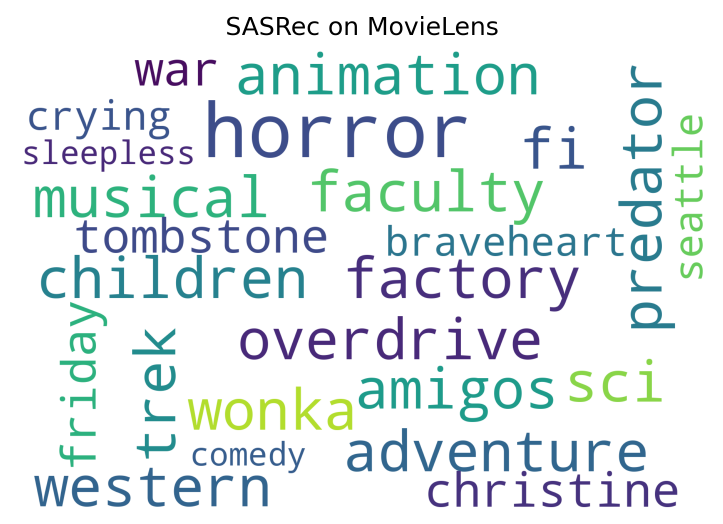}

    \includegraphics[width=0.45\linewidth]{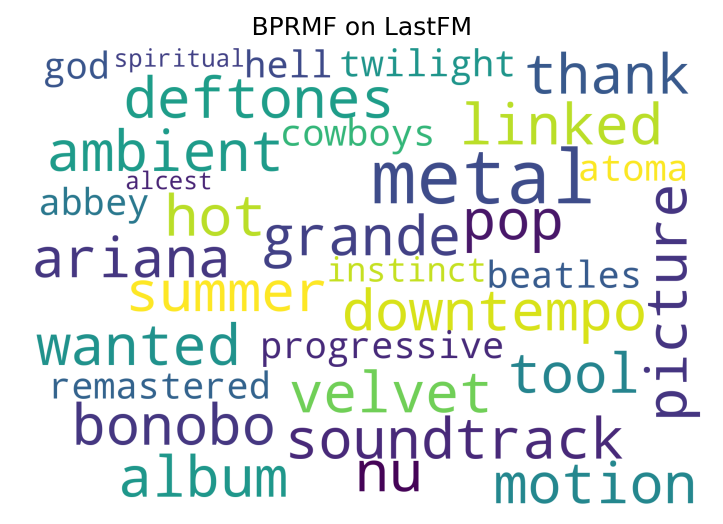}
    \includegraphics[width=0.45\linewidth]{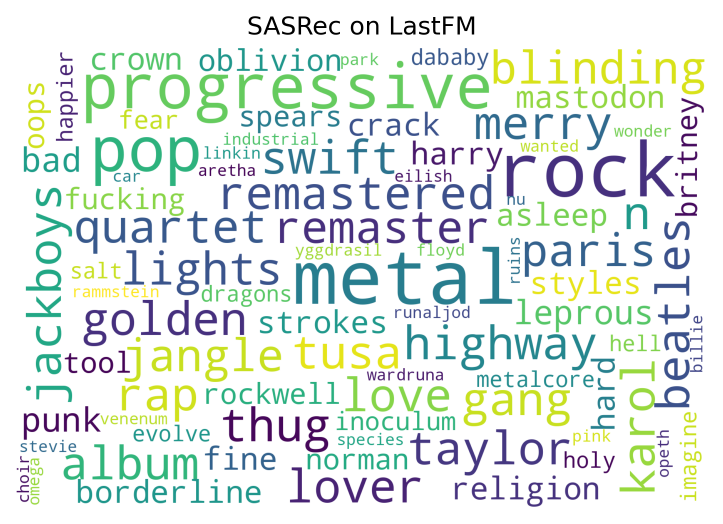}

     \includegraphics[width=0.45\linewidth]{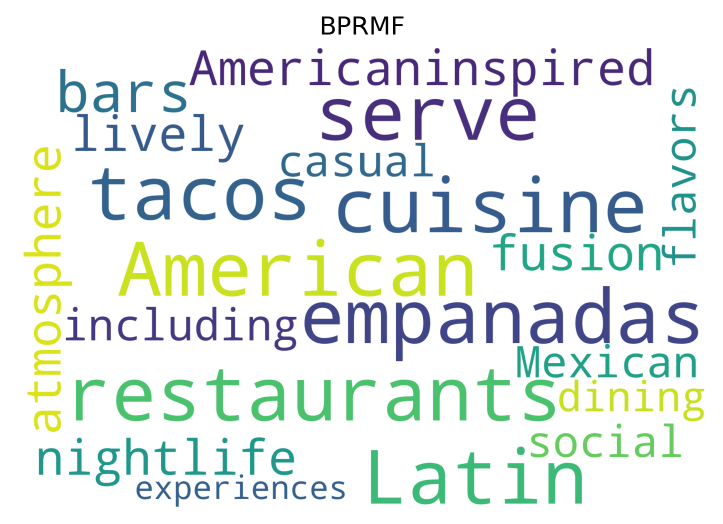}
    \includegraphics[width=0.45\linewidth]{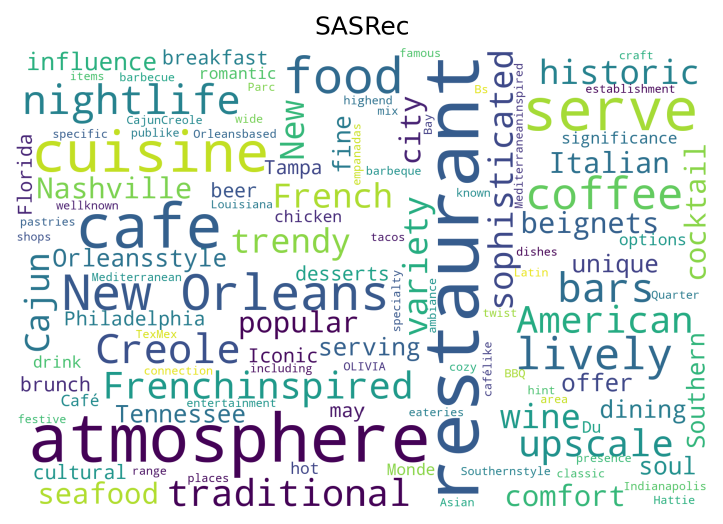}
    
    \caption{Concept Survey on BPRMF~(left) and SASRec~(right) trained on the \textcolor{black}{four} datasets. RecSAE \textcolor{black}{with the LLM interpretation} gains more diverse concepts in models with stronger recommendation performance.}
    \label{fig: word_cloud}
\end{figure}

We perform both qualitative and quantitative analyses of the interpreted concepts. Figure~\ref{fig:word_cloud} presents word clouds generated from the latent concepts of BPRMF and SASRec across \textcolor{black}{four datasets}. Only concepts with a confidence score of 1 are visualized to ensure high reliability. Notably, SASRec captures a broader range of meaningful concepts across all datasets. In the Amazon (Grocery and Gourmet Food) dataset, RecSAE extracts food-related terms such as dietary restrictions ("gluten", "glutino"), types ("cappuccino", "tea", "pasta", "cheeseburger"), packaging/usage features ("microwavable"), and brand names ("Crunchmaster").
In MovieLens, genre-related tags such as "horror", "musical", "children", "comedy", "western", "adventure", as well as specific movie references like "predator", are identified.
In LastFM, musical genres ("metal", "rock", "rap", "pop"), artist names ("Taylor", "Eilish", "Beatles"), album titles ("Jackboys"), and descriptive terms from song titles (e.g., "Remastered") are effectively captured by RecSAE.
\textcolor{black}{In Yelp, business types ("bar", "restaurant"), food types ("seafood", "wine", "brunch", "desserts", "barbecue") and style ("American", "French", "Mexican") are effectively captured by RecSAE.}

Although BPRMF also reveals some interpretable concepts, the breadth and diversity are noticeably more limited compared to the sequential model SASRec. In the Amazon dataset, BPRMF achieves an NDCG@10 of 0.0200, while SASRec attains a higher score of 0.0367. This performance discrepancy is consistent with the richness and coverage of RecSAE's interpreted concepts, highlighting the differences in representational capacity between the two models.

\begin{table}[htbp]
\caption{We quantify the presence of various features (Genre in MovieLens; Music Tag, Astist, Album in LastFM) across interpreted concepts \textcolor{black}{by TF-IDF method} with verification score $\geq$ 0.8. Specifically, we report the number of features that appear in the concepts and the coverage of features.}
    \label{table:concept_survey}
    \centering
    \begin{tabular}{llllll}
    \toprule
  \multirow{2}{*}{\textbf{\makecell[l]{Model}}} & \textbf{MovieLens} & \multicolumn{4}{c}{\textbf{LastFM}} \\
   & Genre & Single-term Tag & All Tag & Artist & Album \\ \hline
      BPRMF   & \textcolor{black}{\textbf{17} (94\%)} & \textcolor{black}{31 (15\%)} &  \textcolor{black}{61 (11\%)} & \textcolor{black}{35 (6\textperthousand)} & \textcolor{black}{125 (10\textperthousand)} \\
      SASRec  & \textcolor{black}{16 (89\%)} & \textcolor{black}{\textbf{55} (27\%)} &  \textcolor{black}{\textbf{109} (20\%)} & \textcolor{black}{\textbf{60} (10\textperthousand)} & \textcolor{black}{\textbf{182} (14\textperthousand)} \\ \bottomrule
    \end{tabular}
\end{table}

Beyond the qualitative analysis, we further quantify the coverage of RecSAE's concepts over the item features provided in the datasets, as reported in Table~\ref{table:concept_survey}. Since the Amazon Grocery and Gourmet Food dataset lacks distinct category labels (items are all labeled as "Grocery and Gourmet Food"), we focus the coverage analysis on the MovieLens (genre) and LastFM (tag, artist, album) datasets. Feature coverage is measured by the ratio of the exact string match between item features and the extracted concept terms to the total number of item features.
In MovieLens, all genre features are successfully identified in SASRec’s concepts. In LastFM, which contains a wider variety of feature types, SASRec’s concepts cover over 20\% of item tags, in contrast to BPRMF, which covers no more than 12\%. This analysis also demonstrates that RecSAE's interpretation results correlate with the model's encoding abilities: higher model accuracy is associated with a greater diversity of interpreted concepts.

% such as "bean", "hazelnut", "vegan", "beef", and "microwaveable". In this dataset, the general model BPRMF achieves a rank of 0.0200 in NDCG@10, while SASRec ranks higher at 0.0367. Meanwhile, RecSAE extracts more diverse concepts from SASRec than from BPRMF, directly revealing the differences in models' encoding abilities. 

% We further show the concept coverage on dataset features in table~\ref{table:concept_survey}.
% As the Amazon Grocery and Gourmet Food does not have item categories

\textbf{\textit{Finding 5: RecSAE can serve as an effective tool for understanding and comparing model encoding behaviors.}}

% \subsubsection{TF-IDF Produced Concepts}

% \subsubsection{LLM Generated Concepts}

\subsection{\textcolor{black}{Concept Quality across Activation Level}}
\label{subsection:add_concept_quality}

\begin{figure}[htbp]
    \centering
    \includegraphics[width=0.45\linewidth]{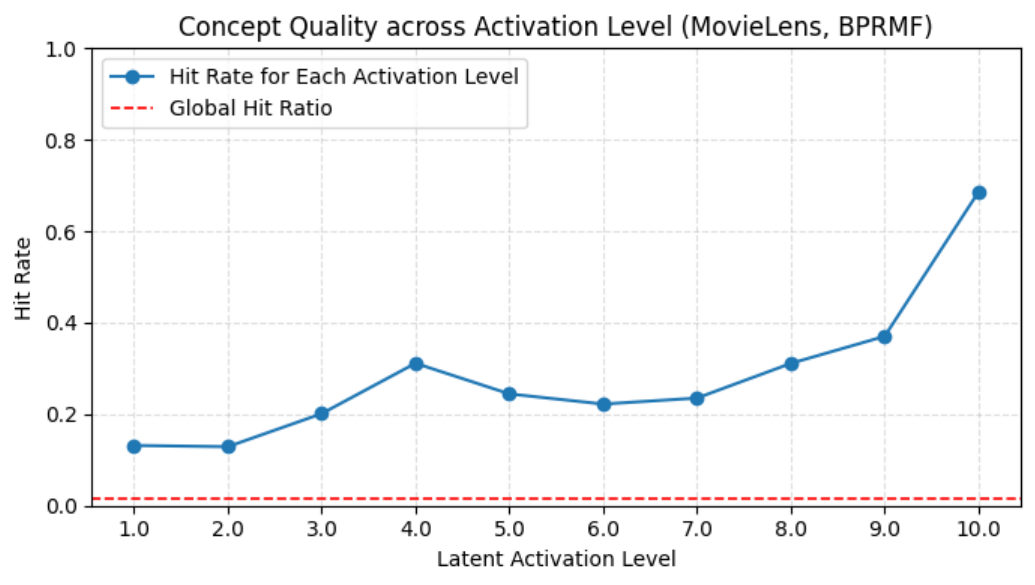}
    \includegraphics[width=0.45\linewidth]{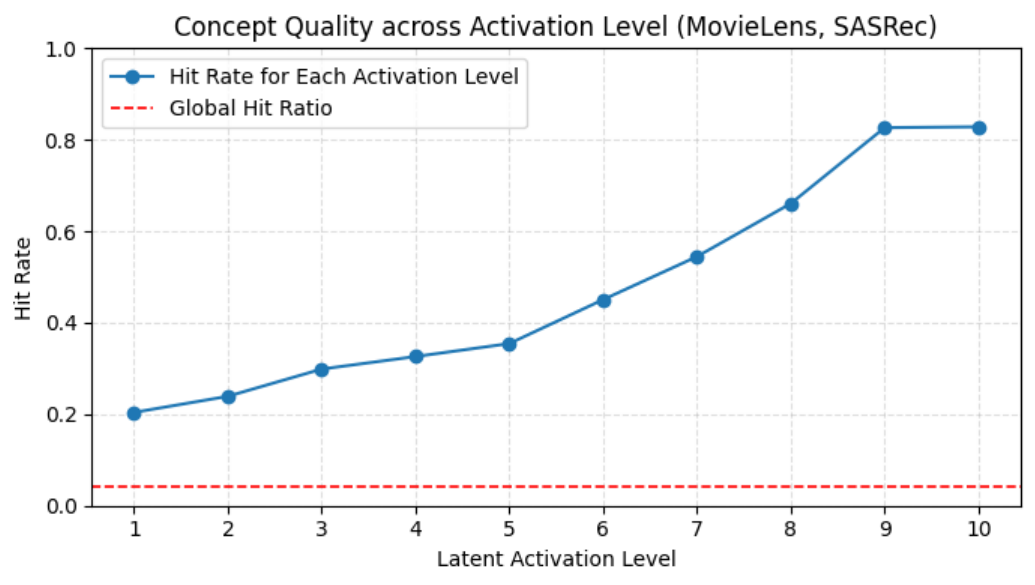}
    \caption{\textcolor{black}{Concept Quality across Latent Activation Levels on the MovieLens Dataset, BPRMF (left) and SASRec (right). When the latent is activated, the corresponding concept hit ratio in the top-1 recommended item (blue line) is substantially higher than the global average (red line).}}
    \label{figure:concept_quality}
\end{figure}

\textcolor{black}{In this section, we further analyze concept quality across different activation levels.
We evaluate the concept hit ratio within recommended items at varying latent activation levels, where the hit ratio measures the presence of the corresponding concept in the top-1 recommended item.
The results are presented in Figure~\ref{figure:concept_quality}.}

\textcolor{black}{The results demonstrate a clear trend: higher activation levels correspond to a stronger presence of the associated concepts in the recommended items (blue line).
Apart from the changes from level 9 to 10 on BPRMF and from level 8 to 9 on SASRec, the overall progression remains smooth, with no distinct cutoff where the concepts become indistinct as the latents lose discriminative power.
Moreover, the average global hit ratios across all features are 0.016 for BPRMF and 0.043 for SASRec (red lines), which are substantially lower than the hit ratios observed when the corresponding latents are activated (blue lines).
This pattern further confirms the strong associations between the latents and their respective concepts.}

\subsection{Human Evaluation of Interpreted Concepts}
\label{subsubsection:human_evaluation}

To Further assess the quality of RecSAE interpretation pipeline, we conduct a human evaluation to determine the relevance of the generated concepts to the items activating the corresponding latents.
Following the annotation protocol established in RecExplainer~\cite{lei2023recexplainer}, three recommendation experts independently annotated 100 randomly selected cases with a latent confidence score above 0.8. A four-level relevance scale was used: 1 - irrelevant, 2 - weakly relevant, 3 - partially similar (some items share similarities), and 4 - highly similar (all items share key similarities).
Specifically, we give annotators the items activating the latent and the corresponding concepts of the latents.

The annotation results show high relevance between latents and corresponding interpreted concepts. \textcolor{black}{We first assess the inter-annotator agreement to evaluate annotation quality. 
Following prior work~\cite{lei2023recexplainer}, we compute the pairwise Cohen’s kappa between every two annotators and report the averaged value, as well as the overall Fleiss’ kappa for multi-annotator agreement.
The agreement scores are 0.3565 for Cohen’s kappa and 0.3407 for Fleiss’ kappa on the 4-level annotations.
These results demonstrate a fair degree of consistency among annotators, comparable to that reported in previous studies on recommendation explanation (e.g., Cohen’s kappa of 0.36 in RecExplainer~\cite{lei2023recexplainer}).}
The average relevance rating across all annotated cases is 3.89 out of 4, indicating that the majority of the interpreted concepts are perceived as highly coherent and semantically meaningful. These findings support the validity of RecSAE’s automated interpretation framework.

% To further examine RecSAE's automatic concept construction and verification pipeline, we conduct human evaluations to assess the relevance of generated concepts with items that activate the latents. Following the human annotation setting established in RecExplainer~\cite{lei2023recexplainer}, three human experts performed four-level relevance annotations (1 irrelevant; 2 relevant; 3 contain similarities among some items; 4 contain key similarities among all items) on randomly selected 100 cases with latent confidence scores greater than 0.8. The agreement between the annotators, measured by Cohen's kappa, is 0.3565, indicating consistency in annotation scores, which is similar to human agreement level in related work in recommendation explanation (e.g. 0.36 in RecExplainer~\cite{lei2023recexplainer}).
% The average relevance rating is 3.890 out of 4, reflecting a high perceived quality of the interpreted concepts. 
% These results validate RecSAE's automatic interpretation pipeline.

% \textbf{\textit{Finding 6: RecSAE interpreted concepts are perceived as highly relevant with latent activated items by human experts.}}

\textbf{\textit{Finding 6: RecSAE interpreted concepts are perceived as relevant (3.89/4) to the items activating the corresponding latents by human experts.}}

\subsection{\textcolor{black}{Comparison with Interpretation Baseline}}
\label{subsection:comparison}

\begin{figure}[htbp]
    \centering
    \includegraphics[width=0.5\linewidth]{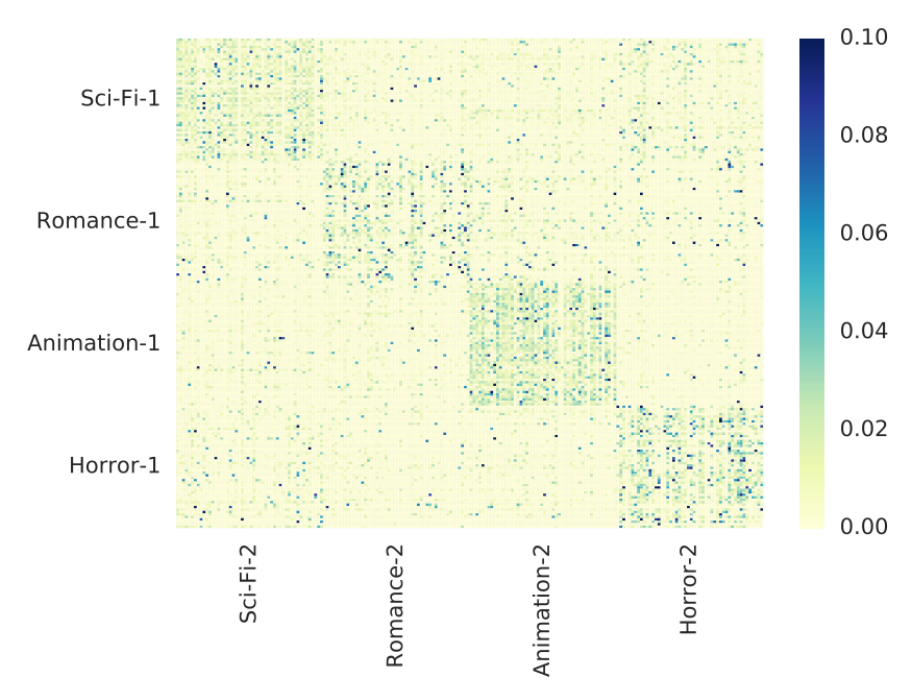}
    \caption{\textcolor{black}{The visualization of SASRec’s attention weights on MovieLens~\cite{kang2018self}. It demonstrates that the attention mechanism can capture items sharing common features. However, it is inherently constrained by the feature set present in the current input sequence, making it difficult to tune the model toward features absent from the current context.}}
    \label{figure:SASRec_case}
\end{figure}

\textcolor{black}{As few methods exist for interpreting different types of recommendation models, we compare the proposed RecSAE framework with baselines specific to different model families. In particular, we evaluate the attention-based explanation of SASRec. 
Following the original SASRec paper, which uses the same model and dataset, we refer to its attention visualization for comparison.  Two disjoint sets of 200 movies are randomly selected from four categories—Science Fiction, Romance, Animation, and Horror—with the first set used as queries and the second as keys. 
Figure~\ref{figure:SASRec_case} illustrates the average attention weights across items, reflecting the model’s ability to capture feature-level similarities.
However, a fundamental limitation of attention-based interpretation and tuning remains: while attention can distinguish feature-level similarities within the observed input, it lacks the capacity to explicitly tune the model’s preference toward unobserved target features. This motivates the need to explicitly model neurons or embeddings specific to each feature beyond attention. }

% \section{RecSAE Reconstruction and Interpretation }
\section{Reconstruction Performance}
\label{section:reconstruction}

In the previous section, we demonstrated the strong interpretability achieved by the RecSAE framework. In this section, we further assess its reconstruction performance to verify that the learned latent space preserves the information contained in the original model’s internal representations.

Table~\ref{table:reconstruction} presents the reconstruction performance. We assess recommendation accuracy using Hit Rate and NDCG by replacing the original model’s internal representations with RecSAE’s reconstructed activations at the designated probing positions. The results indicate that recommendation performance remains largely stable across all four recommendation models and \textcolor{black}{four} datasets, confirming that RecSAE’s latent space effectively preserves the essential information from the original representations.
In comparison to a standard autoencoder (AE) with the same number of latent activations (8), RecSAE’s sparse autoencoder demonstrates significantly better reconstruction performance across all datasets. This highlights the advantage of the sparse autoencoder architecture in retaining critical information from the original model while also encouraging the formation of disentangled, mono-semantic latents that support effective interpretation.

% \todo{other baseline}
% Since the latent space retains only top k activations (k = 8), some performance change may occur. 

It is worth emphasizing that RecSAE does not interfere with the performance of the base recommendation models, as its reconstructions are not used to replace the original representations during the interpretation process.
The RecSAE latents are solely employed for interpretation, and the above reconstruction results show that these latents effectively preserve the information encoded in the original model representations.

% \textbf{\textit{Finding 5: RecSAE reconstructions are }}

\begin{table}[htbp]
\centering
\caption{RecSAE Reconstruction Performance. We evaluate the training with downstream performance when replacing the model's original representations with reconstructions. "N" is NDCG. RecSAE reconstruction performances remain stable, confirming RecSAE latents effectively preserve the essential information from the original model representations.}
\label{table:reconstruction}
\resizebox{0.85\linewidth}{!}{
\begin{tabular}{ll|rr|rr|rr|rr}
\toprule
\multirow{2}{*}{\textbf{\makecell[l]{Rec\\Model}}} & \multirow{2}{*}{\textbf{Method}} 
& \multicolumn{2}{c|}{\textbf{Amazon}} 
& \multicolumn{2}{c|}{\textbf{MovieLens}} 
& \multicolumn{2}{c|}{\textbf{LastFM}} 
& \multicolumn{2}{c}{\textbf{\textcolor{black}{Yelp}}} \\
& & N@10 & N@20 & N@10 & N@20 & N@10 & N@20 & \textcolor{black}{N@10} & \textcolor{black}{N@20} \\ 
\midrule

& Base   & 0.0200 & 0.0248 & 0.0382 & 0.0530 & 0.0702 & 0.0787 & \textcolor{black}{0.0036} & \textcolor{black}{0.0063} \\
BPRMF  & AE   & 0.0157 & 0.0207 & 0.0313 & 0.0429 & 0.0129 & 0.0166 & \textcolor{black}{0.0032} & \textcolor{black}{0.0061} \\
& RecSAE & 0.0192 & 0.0240 & 0.0355 & 0.0499 & 0.0544 & 0.0616 & \textcolor{black}{0.0034} & \textcolor{black}{0.0063} \\
\hline
& Base   & 0.0224 & 0.0284 & 0.0373 & 0.0516 & 0.0952 & 0.1049 & \textcolor{black}{0.0069} & \textcolor{black}{0.0106} \\
LightGCN & AE   & 0.0157 & 0.0204 & 0.0320 & 0.0444 & 0.0127 & 0.0159 & \textcolor{black}{0.0044} & \textcolor{black}{0.0066} \\
& RecSAE & 0.0212 & 0.0268 & 0.0373 & 0.0505 & 0.0755 & 0.0844 & \textcolor{black}{0.0065} & \textcolor{black}{0.0096} \\
\hline
& Base   & 0.0367 & 0.0460 & 0.1049 & 0.1302 & 0.1579 & 0.1656 & \textcolor{black}{0.0043} & \textcolor{black}{0.0065} \\
SASRec  & AE   & 0.0312 & 0.0391 & 0.0441 & 0.0584 & 0.0111 & 0.0146 & \textcolor{black}{0.0040} & \textcolor{black}{0.0062} \\
& RecSAE & 0.0334 & 0.0419 & 0.0973 & 0.1205 & 0.1207 & 0.1288 & \textcolor{black}{0.0040} & \textcolor{black}{0.0060} \\
\hline
& Base   & 0.0355 & 0.0427 & 0.1119 & 0.1369 & 0.1521 & 0.1606 & \textcolor{black}{0.0149} & \textcolor{black}{0.0189} \\
TiMiRec & AE   & 0.0275 & 0.0342 & 0.0490 & 0.0649 & 0.0150 & 0.0195 & \textcolor{black}{0.0054} & \textcolor{black}{0.0076} \\
& RecSAE & 0.0314 & 0.0389 & 0.1083 & 0.1333 & 0.1188 & 0.1271 & \textcolor{black}{0.0119} & \textcolor{black}{0.0160} \\

\bottomrule
\end{tabular}
}
\end{table}

\section{Parameter Sensitivity on Reconstruction and Interpretation}
% \subsection{Reconstruction and Interpretation Frontier with Scale Size s and Sparsity Factor k}
\label{section:recstruct_interpet}

% We conduct a series of experiments to examine the impact of hyper-parameters.
% Scale size $s$ determines latent space dimensions and sparsity factor $k$ controls the sparsity level of latent activations.

We conduct sensitivity experiments to examine the influence of hyper-parameters in the RecSAE framework. Specifically, we evaluate how the scale size $s$, which determines the dimensionality of the latent space, and the sparsity factor $k$, which controls the number of active latents per instance, affect both the reconstruction quality and the interpretability of the learned representations.

\subsection{Scale Size $s$}
\begin{figure}[htbp]
    \centering
    \includegraphics[width=0.95\linewidth]{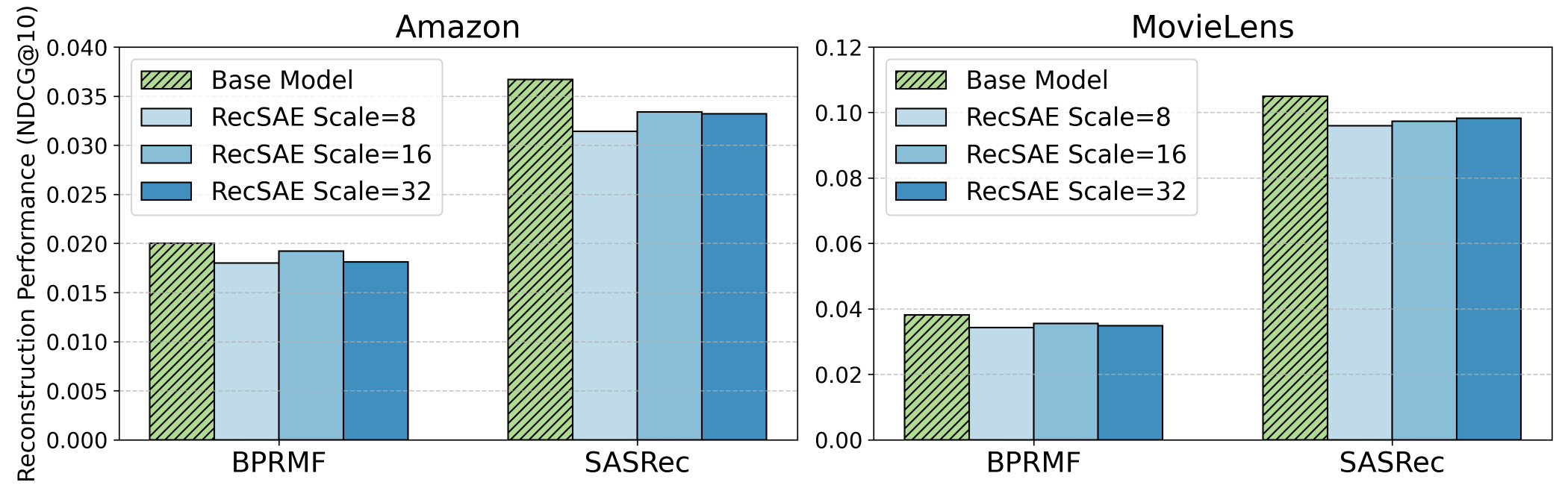}
    \caption{RecSAE achieves optimal reconstructions when the scale size $s$=16, corresponding to 1024 latents (16* 64 base model dimensions). This trend is consistent for both sparsity factor $k$=8 and $k$=16; in the figure, results are shown for $k$=8.}
    \label{fig:parameter_scale_size}
\end{figure}
% We consider the reconstruction performance when choosing RecSAE scale sizes. 
% We first fix the sparsity factor $k$=8 and vary $s$. 
% As shown in Figure~\ref{fig:parameter_scale_size},  reconstruction performance peaks when $s$=16, corresponding to 1024 latent dimensions.
% This suggests that 1024 latents are sufficient to encode the 64-dimensional dense internal representations of the investigated recommendation models.
% This pattern holds for both $k$=8 and $k$=16. Based on these observations, we set $s$=16 in experiments.

We first investigate the impact of varying the scale size $s$ on reconstruction performance. In this setting, we fix the sparsity factor at $k = 8$ and vary $s$ to explore how the dimensionality of the latent space influences RecSAE’s ability to reconstruct the internal representations of recommendation models.

As shown in Figure~\ref{fig:parameter_scale_size}, in most settings (3 out of 4), the reconstruction performance peaks at $s = 16$, which corresponds to 1024 latent dimensions. Deviating from this configuration, either by reducing or increasing the number of latent dimensions, generally results in a degradation of reconstruction quality. This finding suggests that a latent space of moderate size is sufficient to capture the essential information encoded in the 64-dimensional dense representations output by the investigated models. 
% We observe that this pattern remains consistent across different values of $k$, including both $k = 8$ and $k = 16$, indicating robustness to sparsity settings. 
Based on these results, we adopt $s = 16$ as the default configuration in our experiments.

% reconstruction performance consistently improves with increasing $s$, peaking at $s = 16$, which corresponds to 1024 latent dimensions. This finding suggests that a latent space of moderate size is sufficient to capture the essential information encoded in the 64-dimensional dense representations output by the investigated models. Importantly, we observe that this pattern remains consistent across different values of $k$, including both $k = 8$ and $k = 16$, indicating robustness to sparsity settings. Based on these results, we adopt $s = 16$ as the default configuration in our experiments.

\begin{figure}[htbp]
    \centering

     \includegraphics[width=0.49\linewidth]{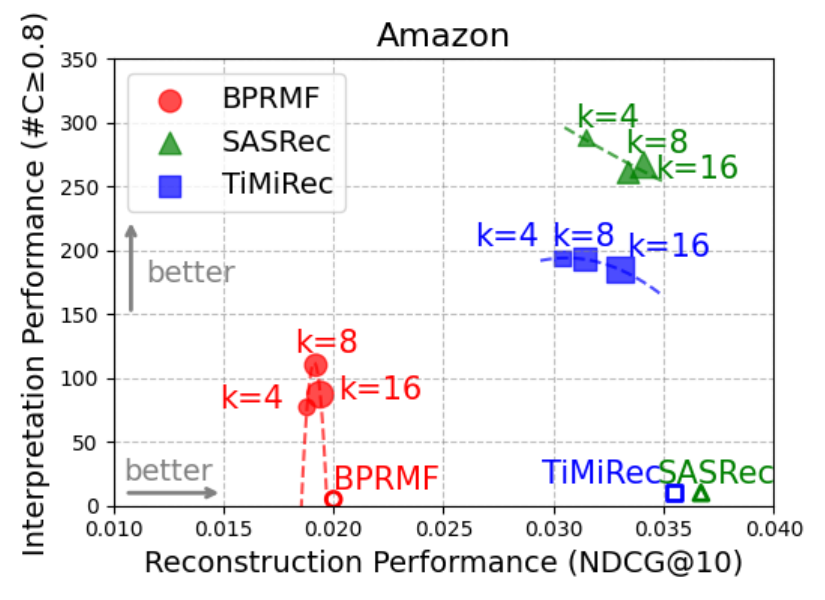}
    \includegraphics[width=0.49\linewidth]{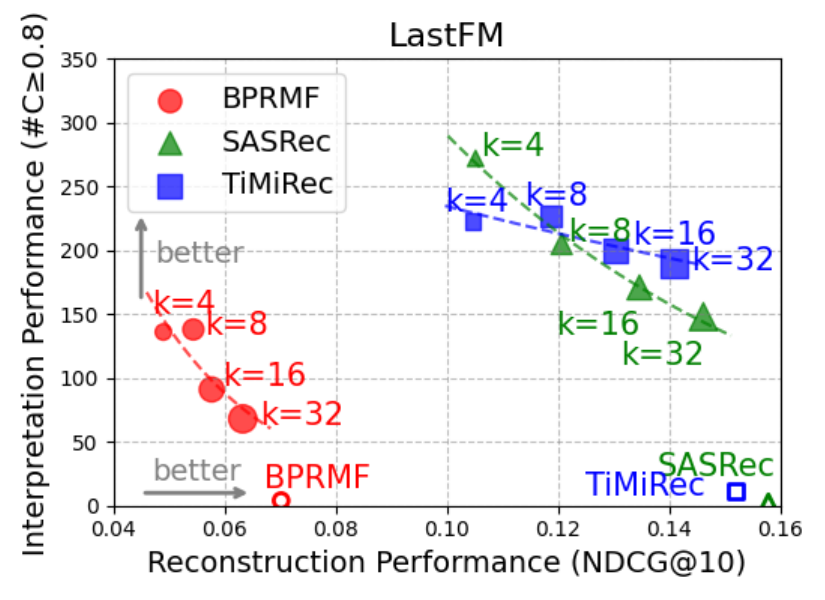}
    \caption{\textcolor{black}{Impact of Sparsity Factor k on Reconstruction and Interpretation. Note that RecSAE does not affect model performances as the reconstructions do not need to be placed into original models during interpretations.}}
    \label{fig:scaling}
\end{figure}

\subsection{Sparsity Factor $k$}
\label{subsubsection:k}
$k$ controls the number of active latents in the latent space, impacting both the reconstruction quality and the interpretability of the RecSAE framework. \textcolor{black}{Figure~\ref{fig:scaling}} illustrates the trade-off of the two factors. LightGCN is not included in the figure as its pattern is similar to BPRMF.
It is important to note that RecSAE does not affect model performance, as the reconstructions do not need to replace the original model activations. Instead, it measures training performance of RecSAE to ensure that the latent space encodes the main information of the interpreted model.
As shown in the figure, increasing $k$ typically improves reconstruction performance but results in fewer concepts with high interpretation confidence.
As \(k\) increases, each item/user is encoded by more latents, leading to a more redundant and less semantically distinct latent space. However, in two cases (BPRMF and TiMiRec on Amazon), the curve peaks at \(k=8\), likely because smaller values of \(k\) may lead to a relatively small number of activations, causing some latents to be insufficiently explained.
In experiments, we set $k$=8 as a balance point between reconstruction quality and interpretability.

% \subsection{Top-N Recommendation Items}
% \label{subsection:add_topN}

% \textcolor{black}{We further evaluate the recommendation Top-N effect on interpretation.}

% \begin{figure}
%     \centering
%     \includegraphics[width=0.48\linewidth]{figures/recommendation_n/topk_amazon_eval=1_all.png}
%     \includegraphics[width=0.48\linewidth]{figures/recommendation_n/topk_movielens_eval=1_all.png}
%     %  \includegraphics[width=0.48\linewidth]{figures/recommendation_n/topk_amazon_eval=1_confidence>=08.png}
%     % \includegraphics[width=0.48\linewidth]{figures/recommendation_n/topk_movielens_eval=1_confidence>=08.png}
%     \caption{\textcolor{black}{Recommendation Top-N's effect on RecSAE Interpretation. We change N from 1 to 1 on four recommendation models on Amazon (left) and MovieLens (right) dataset and show the interpreted concept numbers.}}
%     \label{figure:recommendationN}
% \end{figure}

\section{Targeted Tuning}
\label{section:targeted_debiasing}

In the preceding section, we demonstrated the efficacy of RecSAE in interpreting various types of recommendation models. Building upon these results, this section explores the potential of RecSAE in enabling targeted tuning of model behavior based on the interpretation outcomes. 
Specifically, the tuning is achieved by modifying the latent activations, which are then reconstructed and used to replace the original recommendation representations. As a result, the modified reconstructions either amplify or attenuate the targeted information.

We begin by presenting case studies where we vary the activation scale of latent variables. This variation leads to monotonic trends in the concept presence ratio among recommended items. Subsequently, we quantitatively report the impact of altering each latent variable and average concept presence ratios, thereby providing evaluations of the overall performance.

Formally, the concept presence ratio is calculated as follows, where $\text{Test}$ represents the entire test set, $\mathbf{S}_l^{(t)}$ denotes the activation value of latent $l$ for input case $t$, $\mathbf{C}_l$ represents the interpreted concept associated with latent $l$, and $I_{pred}^{(t)}$ is the predicted item list (Top 10 items in later evaluations) for input case $t$. The indicator function $\mathbb{I}(\cdot)$ is used to evaluate whether a given condition holds true.

\begin{equation}
\label{equation:ratio}
\begin{split}
   \text{Presence Ratio}(l) = \frac{\sum_{t}^{\text{Test}} \mathbb{I}\left( \mathbf{S}_l^{(t)} \neq 0 \land \mathbf{C}_l \in i_{pred}^{(t)} \right)}{\sum_{t}^{\text{Test}} \mathbb{I}\left( \mathbf{S}_j^{(t)} \neq 0 \right)}
   % \text{Presence Ratio}(l) = \frac{\sum\_{t \in \text{Test}} \mathbb{I}\left( \mathbf{S}*l^{(t)} \neq 0 \land \mathbf{C}*l \in I*{pred}^{(t)} \right)}{\sum*{t \in \text{Test}} \mathbb{I}\left( \mathbf{S}\_l^{(t)} \neq 0 \right)}
\end{split}
\end{equation}

\subsection{Case Study}

As shown in Table~\ref{table:concept_survey}, RecSAE maps distinct item features to separate latents, demonstrating the model’s ability to isolate and represent individual features. Building on this, we identify these latents for targeted tuning. To explore this capability, we manipulate the activation of a single latent while keeping all others fixed, and observe the resulting changes in the model’s recommendation outputs.

\begin{figure}[htbp]
    \centering
    \includegraphics[width=0.4\linewidth]{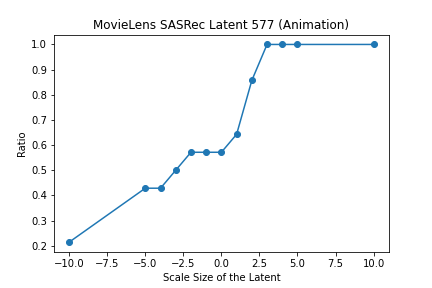}
    \includegraphics[width=0.4\linewidth]{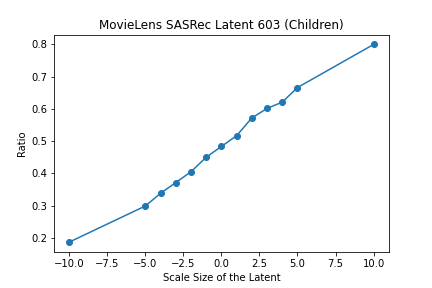}
    \includegraphics[width=0.4\linewidth]{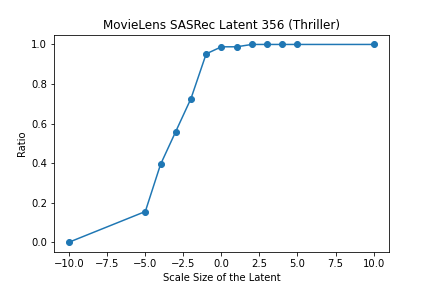}
    \includegraphics[width=0.4\linewidth]{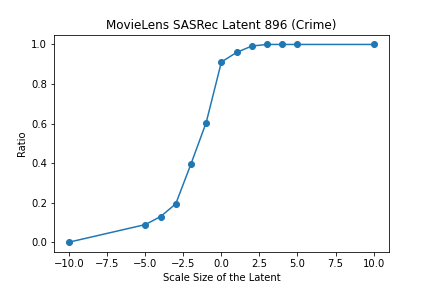}
    \caption{By scaling the activation of each latent and measuring the resulting concept presence ratio in the recommended items, we demonstrate that this targeted modulation effectively steers the model’s outputs toward the corresponding direction.}
    \label{fig:ablation}
\end{figure}

In this process, the original activation values, denoted as scale 1, are systematically replaced with a set of scaled values: -10, -5 to 5 in steps of 1, and 10.
For each of these scales, we obtain the corresponding recommendation outputs of the model, and compute the concept Presence Ratio, as defined in Equation~\ref{equation:ratio}. The results, presented in Figure~\ref{fig:ablation}, reveal a clear trend: increasing the activation values of latents leads to a higher ratio of items related to the target concept in the recommendations, while decreasing the activation values results in a reduction in the presence of such items. These monotonic trends illustrate RecSAE’s capacity for targeted tuning, offering a robust mechanism for adjusting model outputs based on specific latent activations.

In practice, this mechanism enables the amplification of particular concepts, such as child-friendly categories (e.g., “children’s,” “animation”), or the attenuation of undesirable content for children (e.g., “crime,” “thriller”) in the model’s outputs. This provides a powerful tool for targeted tuning model outputs to align with certain desired preferences and requirements.

\subsection{Quantitative Report of Tuning Performance}

The case study presented above demonstrates the monotonic relationship between latent activation scales and the corresponding concept presence in the output items. Building on these observations, we conduct further quantitative evaluations to assess the overall performance of tuning each latent. Specifically, we adjust the activation levels of each latent from a -10× scale to a +10× scale, obtaining the corresponding output recommendations. For each latent, we calculate the presence ratios of both the related and other concepts, reporting the averages across all latents. The presence ratio of the related concept reflects the efficacy of the tuning, and the presence ratio of the other latents serves as an indicator of the tuning precision, ensuring that unrelated concepts are minimally impacted.

The results are reported in Table~\ref{table:ablation}. Across the three datasets, the monotonic trend of related concept persists: decreasing the activation values of the latents results in a lower concept presence ratio, while increasing the activation values leads to a higher ratio. The average variance in concept presence exceeds 10\%. Although increasing the latent activations does not result in a 100\% presence ratio, this outcome is understandable, given that there are eight latents activated in each cases and the modified latents may not dominate the overall behavior. If the original activation value of a latent is relatively low compared to others, scaling it may not always lead to the related concepts ranked in the top 10 item list.

Regarding the unrelated concepts, modifications across the three settings (decreased~(-10×), original~(+1×), and increased~(+10×)) lead to minimal changes in outputs. s the rank of related items increases, some minor changes in the ratio of unrelated items are observed; however, these shifts remain minimal, not exceeding 0.5\%. This is significantly lower than the changes observed in the related concepts. The relatively small variation in the ratio of unrelated items, in contrast to the more substantial shifts in related item presence, highlights the precision and effectiveness of RecSAE’s targeted tuning.

% As for the other unrelated concepts, the modification across decreased~(-10×), original~(+1×) and increased~(+10×) leads to minor change in outputs. Although there are changes, it is no more than 0.6\% percentage. As the increase of related items raise, there are reasonable change in the ratio of unrelated items. This minor change compared to the much larger related item changes show the precision of the targeted tuning of RecSAE.

In conclusion, RecSAE demonstrates its potential as a powerful tool for targeted tuning of model behavior. By carefully adjusting the activation values of relevant and irrelevant latents, and applying these tuned reconstructions, recommendation models can be effectively customized for specific tasks and requirements.

% The above case study show the example latents' mono-trend between latent activation scale and corresponding concept presence in output items.
% Beyond these cases, we further conduct quantitative evaluations to examine the overall situations while tuning each latent.
% We tune each latent' activation level from 1 to -10 scale and +10 scale, getting the corresponding output and calculate each latents' concept presence ratio and average the ratio across latent.

% The results are reported in Table~\ref{table:ablation}. Across the three datasets, the mono-trend holds. Decreasing the latents leads to low concept presence ratio in more than and increase the latent activation values raise the value. The average variance in more than 10 percentages. Although increasing the latent activations does not leads to 100\% percentage, it is understandable that there are eight latents and the modifying cases might not be the dominate latents. If its original activation value is low compared to other activated cases, it is reasonable that scale it 10 times do not always leads to the related concepts in output items.
% In overall situation, RecSAE serve as a promising tool for targeted tuning model beharior.
% By revising and desired and undesired latents with tuned reconstructions, recommenders can be targeted tuned.

\begin{table}[htbp]
 \caption{Targeted Tuning to SASRec trained on the three dataset. we modify the activation value of each latent by scaling it to -10× and +10×, and report the presence ratios for Related Concept and Other Concepts under the three settings: Decreased~(-10×), Original~(+1×), and Increased~(+10×). The high variation in related concepts and the minimal change in other concepts demonstrate the efficacy of RecSAE's targeted tuning capabilities.}
    \label{table:ablation}
    \centering
    \begin{tabular}{lrrrrrr}
    \toprule
        % \multirow{2}{*}{Dataset}  & \multicolumn{2}{c}{Decreased~(-10×)} & \multicolumn{2}{c}{Original~(+1×)} & \multicolumn{2}{c}{Increased~(+10×)}   \\
        % & Related & Others & Related & Others & Related & Others \\ \midrule
        % Amazon & 9.3\%  & 6.5\% & 30.1\% & 7.1\% & 43.2\%  & 6.8\% \\
        % % MovieLens & 3.3\% & 30.6\% & 39.2\% \\
        % MovieLens & 3.8\% & 2.4\% & 35.7\% & 2.9\% & 47.5\%  &  2.7\% \\
        % LastFM & 11.2\%  & 4.3\%  & 23.3\% & 4.0\%  & 32.3\% & 4.0\%  \\ 

         \multirow{2}{*}{\textbf{Dataset}}  & \multicolumn{3}{c}{\textbf{Related Concept}} & \multicolumn{3}{c}{\textbf{Other Concepts}} \\
        & Decreased & Original & Increased & Decreased & Original & Increased \\ \midrule
        Amazon & 9.3\%  & 30.1\% &  43.2\% & 6.7\% & 7.2\% & 6.9\%  \\
        % Amazon & 9.3\%  & 6.5\% & 30.1\% & 7.1\% & 43.2\%  & 6.8\% \\
        MovieLens & 3.8\% & 35.7\% & 47.5\% & 2.4\% & 2.9\%  &  2.7\%\\
        % 2.3 2.8 2.7
        % MovieLens & 3.8\% & 2.4\% & 35.7\% & 2.9\% & 47.5\%  &  2.7\% \\
        LastFM & 11.2\%  & 23.3\%  & 32.3\% & 4.3\% & 4.0\% & 4.0\%   \\ 
        % LastFM & 11.2\%  & 4.3\%  & 23.3\% & 4.0\%  & 32.3\% & 4.0\%  \\ 

        \bottomrule
    \end{tabular}
\end{table}

% The RecSAE framework has the potential to make predictable changes to model behavior. Furthermore, RecSAE may uncover concepts related to popularity bias or unfairness toward users and items. By revising these "harmful" latents with tuned fairer reconstructions, recommenders can be effectively de-biased.

% \begin{table}[htbp]
%     \centering
%     \caption{Targeted Tuning. We scale latent value and examine whether the concept appears in top 10 recommended items.}
%     \label{table:debiasing}
%     % \resizebox{0.88\linewidth}{!}{
%     \begin{tabular}{lccccc}
%     \toprule
%        \multirow{2}{*}{Dataset} & \multirow{2}{*}{Latent ID} & \multirow{2}{*}{Concept} &  \multicolumn{3}{c}{Hit@10}  \\ 
%          &  &  & -10× & Original & +10× \\ \hline
%        Amazon & 288 & Coconut & 0.0189 & 0.6604 & 0.9528  \\
%         MovieLens & 463 & Horror & 0.0000 & 0.9213 & 1.0000\\ 
%        LastFM & 67 & Taylor Swift  & 0.0000 & 0.4079 & 1.0000\\ 
       
%        \bottomrule
%     \end{tabular} 
%     % }
% \end{table}

\textit{\textbf{Finding 7: RecSAE shows the potential for targeted tuning of model behaviors.}}

% 271
% 684
% 804

\section{Discussion and Future work}
\label{sec:discussion}

\subsection{Complexity}
\label{subsection:complexity}
Assuming the probing site in the original model has a dimension of $|D|$ (typically less than hundreds in recommenders), the space complexity is \( O(s|D|^2) \), where $s$ is the scale size of latent space (in our experiments the optimal $s$ is 16). Training and inference have a time complexity of \( O(s|D|^2)\). 
As recommendation models typically do not scale with embedding dimensions~\cite{sethi2022recshard}, and the complexity of RecSAE depends only on the scale size and embedding dimensions of the interpreted model, rather than the overall complexity of the base recommender, the RecSAE framework holds the potential for interpreting industry-level recommendation systems.

\subsection{Future Work}
\label{subsection:future_work}

This paper presents a structured design and evaluation of RecSAE's capabilities. While RecSAE introduces a new paradigm for interpreting and tuning various recommendation models, several aspects remain open for further exploration. The key areas include:

% This paper provides a structured design and evaluation of RecSAE's abilities.
% While as RecSAE opens a new paradigm for interpreting and tuning various recommendation models, there are many aspects that remained further explorations. The major aspects are:

1) Generalize to \textbf{more types of recommendation models} 
% , such as content-based models and industry-level systems.

RecSAE's framework is highly flexible and scalable. The probing paradigm is adaptable, and the interpretation process is automated. These features make it suitable for integration with various types of recommendation models, including general, graph-based, context-based, sequential, content-based models and large-scale industry systems. \textcolor{black}{Besides, industrial-scale recommendation systems often involve complex multi-stage pipelines (retrieval, ranking, and re-ranking).
RecSAE can be integrated into such settings to evaluate how different stages capture varying levels of feature granularity—where the retrieval stage tends to model coarse-grained features, while finer-grained and more personalized features are progressively incorporated in the ranking and re-ranking stages.} Future work could extend RecSAE to these additional models and stages, providing interpretations and comparable analysis to various type of systems.

% RecSAE framework is highly flexible and scalable. Its probing paradigm can be integrated with various types of recommendation models and the interpretation process is automated. This adaptability and efficiency enable RecSAE to interpret a wide range of recommendation models.

2) Investigate the impact of \textbf{different probing positions} across model layers
% to better understand information processing dynamics.

\textcolor{black}{As user embeddings are present in almost all recommendation models and serve as the core representation for personalized modeling, we probe this layer to evaluate the generalizability and effectiveness of RecSAE across diverse backbones. We acknowledge that model-aware probing at deeper layers reflecting the model’s internal reasoning process would provide complementary insights.}
% In this study, we probe at the user representations, which are typically presented in recommendation models.
Future research could explore the impact of probing at different positions within the model layers, based on the specific architecture of each model. This could provide deeper insights into how information is processed across layers and help researchers and system developers better understand the internal workflows of recommendation models. Additionally, investigations could also examine the concept formulation dynamics across model training process. This could reveal how the model progressively learns different levels of concepts as training evolves.
% In this paper, we adopt the user representations as the probing position. This representation is presented in each recommendation models and usually near the end layers. Future analysis could probing at positions according to each model structures and investigate the concept formulation across model layers. The model-specific designs could help research and system developers to better understanding the model's internal workflows. Besides, RecSAE could also help to investigate the training dynamics. We can see that as the training process develop, how model learns different level of concepts.

\textcolor{black}{3) Establish more \textbf{fine-grained and in-depth training and interpretation pipelines}\\
The current two-step paradigm, comprising unsupervised training followed by post-hoc interpretation, offers certain advantages. It allows concepts to emerge objectively from the learned latents, without relying on a pre-defined concept set that could introduce bias during alignment training.
Future directions could also explore a supervised method during training with alignment losses for better interpretation performance.
For interpretation, the current pipeline consists of two components: concept construction and concept verification. It also employs two tools: a TF-IDF-based method for lightweight concept extraction and an LLM for deeper semantic understanding.
While the current designs are carefully established, future work could explore more fine-grained pipelines, such as incorporating n-gram features into the TF-IDF stage, penalizing terms appearing in negative samples during construction, utilizing stronger LLMs beyond the current 8B open-source model, and developing more robust verification procedures. 
Moreover, future work could incorporate uncertainty-aware interpretation methods, such as selecting items based on a lower confidence bound to identify those most strongly associated with each latent’s semantics, rather than directly relying on the Top-N recommended items for the most activated cases.
These improvements could further enhance the quality of the interpreted concepts derived from the learned sparse latents.
}

4) Conduct detailed evaluations of RecSAE’s utility for \textbf{targeted tuning and de-biasing}
% to improve model fairness.

In this paper, we demonstrate the potential of RecSAE for targeted tuning of model behavior. Specifically, Section~\ref{section:targeted_debiasing} shows how RecSAE can influence the presence of interpreted concepts in model outputs by modifying latent activations. Further developments could focus on using RecSAE for targeted debiasing to enhance model performance and fairness. For example, RecSAE could help reduce the sensitive model outputs to specific users, such as child-unfriendly categories, and promote more diverse or niche content. Additionally, RecSAE could be used to mitigate popularity bias in recommendation models, improving fairness. These capabilities highlight the promising potential of RecSAE to address issues and improve performance in recommendation systems.

% In this paper, we show the great potential for conduct targeted tuning to model behavior with RecSAE framework. The section~\ref{section:targeted_debiasing} shows that RecSAE is able to increase or decrease interpreted concepts in model output by varying the latent activations. Further designs could help to remove sensitivity content to certain groups (child-unfriendly), help to push niche or diverse contents to improve fairness and conduct popularity de-bias to model behaviors.
% This are the reachable potential benefit of RecSAE.

\section{Conclusion}
\label{sec:conclusion}
In this work, we propose the RecSAE framework with a probing paradigm for interpreting various types of recommendation models. It extracts interpretable latents from models' internal activations and links these latents to semantic concepts, which are further validated to obtain confidence scores. These concepts are human-readable and easy to verify. The RecSAE framework represents a novel step toward interpreting models without compromising their performances, while also enabling targeted tuning to models based on the interpretation results. Experiments conducted on three types of recommendation models across widely used datasets demonstrate the effectiveness and generalizability of the RecSAE framework.

\section*{Acknowledge}
% This work is supported by the Natural Science Foundation of China (Grant No. U21B2026) and Quan Cheng Laboratory
% (Grant No QCLZD202301).
This work is supported by the Natural Science Foundation of China (Grant No. U21B2026, 62372260), the National Key Research and Development Program of China under Grant 2024YFC3307403, and Research Project of Quan Cheng Laboratory, China (Grant No. QCL20250105).

\bibliographystyle{ACM-Reference-Format}
\bibliography{sample-base}


\begin{thebibliography}{62}

%%% the Web address.

\ifx \showCODEN    \undefined \def \showCODEN     #1{\unskip}     \fi
\ifx \showISBNx    \undefined \def \showISBNx     #1{\unskip}     \fi
\ifx \showISBNxiii \undefined \def \showISBNxiii  #1{\unskip}     \fi
\ifx \showISSN     \undefined \def \showISSN      #1{\unskip}     \fi
\ifx \showLCCN     \undefined \def \showLCCN      #1{\unskip}     \fi
\ifx \shownote     \undefined \def \shownote      #1{#1}          \fi
\ifx \showarticletitle \undefined \def \showarticletitle #1{#1}   \fi
\ifx \showURL      \undefined \def \showURL       {\relax}        \fi
% invisible to TeX
\providecommand\bibfield[2]{#2}
\providecommand\bibinfo[2]{#2}
\providecommand\natexlab[1]{#1}
\providecommand\showeprint[2][]{arXiv:#2}

\bibitem[AI@Meta(2024)]%
        {llama3modelcard}
\bibfield{author}{\bibinfo{person}{AI@Meta}.} \bibinfo{year}{2024}\natexlab{}.
\newblock \showarticletitle{Llama 3 Model Card}.
\newblock  (\bibinfo{year}{2024}).
\newblock
\urldef\tempurl%
\url{https://github.com/meta-llama/llama3/blob/main/MODEL_CARD.md}
\showURL{%
\tempurl}


\bibitem[Ann.(2025)]%
        {sae_limitation_iclr}
\bibfield{author}{\bibinfo{person}{Ann.}} \bibinfo{year}{2025}\natexlab{}.
\newblock \showarticletitle{On the Limits of Sparse Autoencoders: A Theoretical Framework and Reweighted Remedy}.
\newblock
\urldef\tempurl%
\url{https://openreview.net/forum?id=DSOTgzeH3w&utm_source=chatgpt.com}
\showURL{%
\tempurl}


\bibitem[Anthropic(2024)]%
        {sae_limitation_anthropic}
\bibfield{author}{\bibinfo{person}{Anthropic}.} \bibinfo{year}{2024}\natexlab{}.
\newblock \showarticletitle{The engineering challenges of scaling interpretability}.
\newblock
\urldef\tempurl%
\url{https://www.anthropic.com/research/engineering-challenges-interpretability?utm_source=chatgpt.com}
\showURL{%
\tempurl}


\bibitem[Balog et~al\mbox{.}(2019)]%
        {balog2019transparent}
\bibfield{author}{\bibinfo{person}{Krisztian Balog}, \bibinfo{person}{Filip Radlinski}, {and} \bibinfo{person}{Shushan Arakelyan}.} \bibinfo{year}{2019}\natexlab{}.
\newblock \showarticletitle{Transparent, scrutable and explainable user models for personalized recommendation}. In \bibinfo{booktitle}{\emph{Proceedings of the 42nd international acm sigir conference on research and development in information retrieval}}. \bibinfo{pages}{265--274}.
\newblock


\bibitem[Bills et~al\mbox{.}(2023)]%
        {openai_exp_neuron}
\bibfield{author}{\bibinfo{person}{Steven Bills}, \bibinfo{person}{Nick Cammarata}, \bibinfo{person}{Dan Mossing}, \bibinfo{person}{Henk Tillman}, \bibinfo{person}{Leo Gao}, \bibinfo{person}{Gabriel Goh}, \bibinfo{person}{Ilya Sutskever}, \bibinfo{person}{Jan Leike}, \bibinfo{person}{Jeff Wu}, {and} \bibinfo{person}{William Saunders}.} \bibinfo{year}{2023}\natexlab{}.
\newblock \showarticletitle{Language models can explain neurons in language models}.
\newblock \bibinfo{journal}{\emph{OpenAI}} (\bibinfo{year}{2023}).
\newblock


\bibitem[Bricken et~al\mbox{.}(2023)]%
        {claude_towards}
\bibfield{author}{\bibinfo{person}{Trenton Bricken}, \bibinfo{person}{Adly Templeton}, {and} \bibinfo{person}{Others}.} \bibinfo{year}{2023}\natexlab{}.
\newblock \showarticletitle{Towards Monosemanticity: Decomposing Language Models With Dictionary Learning}.
\newblock \bibinfo{journal}{\emph{Anthropic}} (\bibinfo{year}{2023}).
\newblock


\bibitem[Christian et~al\mbox{.}(2016)]%
        {christian2016single}
\bibfield{author}{\bibinfo{person}{Hans Christian}, \bibinfo{person}{Mikhael~Pramodana Agus}, {and} \bibinfo{person}{Derwin Suhartono}.} \bibinfo{year}{2016}\natexlab{}.
\newblock \showarticletitle{Single document automatic text summarization using term frequency-inverse document frequency (TF-IDF)}.
\newblock \bibinfo{journal}{\emph{ComTech: Computer, Mathematics and Engineering Applications}} \bibinfo{volume}{7}, \bibinfo{number}{4} (\bibinfo{year}{2016}), \bibinfo{pages}{285--294}.
\newblock


\bibitem[Cunningham et~al\mbox{.}(2023)]%
        {cunningham2023sparse}
\bibfield{author}{\bibinfo{person}{Hoagy Cunningham}, \bibinfo{person}{Aidan Ewart}, \bibinfo{person}{Logan Riggs}, \bibinfo{person}{Robert Huben}, {and} \bibinfo{person}{Lee Sharkey}.} \bibinfo{year}{2023}\natexlab{}.
\newblock \showarticletitle{Sparse autoencoders find highly interpretable features in language models}.
\newblock \bibinfo{journal}{\emph{arXiv preprint arXiv:2309.08600}} (\bibinfo{year}{2023}).
\newblock


\bibitem[Du et~al\mbox{.}(2023)]%
        {ClusteringDu2023}
\bibfield{author}{\bibinfo{person}{Yuntao Du}, \bibinfo{person}{Jianxun Lian}, \bibinfo{person}{Jing Yao}, \bibinfo{person}{Xiting Wang}, \bibinfo{person}{Mingqi Wu}, \bibinfo{person}{Lu Chen}, \bibinfo{person}{Yunjun Gao}, {and} \bibinfo{person}{Xing Xie}.} \bibinfo{year}{2023}\natexlab{}.
\newblock \showarticletitle{Towards Explainable Collaborative Filtering with Taste Clusters Learning}. In \bibinfo{booktitle}{\emph{Proceedings of the ACM Web Conference 2023}}. \bibinfo{publisher}{ACM}, \bibinfo{pages}{3712–3722}.
\newblock
\href{https://doi.org/10.1145/3543507.3583303}{doi:\nolinkurl{10.1145/3543507.3583303}}


\bibitem[Du et~al\mbox{.}(2024)]%
        {du2024tutorial}
\bibfield{author}{\bibinfo{person}{Zhaocheng Du}, \bibinfo{person}{Chuhan Wu}, \bibinfo{person}{Qinglin Jia}, \bibinfo{person}{Jieming Zhu}, {and} \bibinfo{person}{Xu Chen}.} \bibinfo{year}{2024}\natexlab{}.
\newblock \showarticletitle{A Tutorial on Feature Interpretation in Recommender Systems}. In \bibinfo{booktitle}{\emph{Proceedings of the 18th ACM Conference on Recommender Systems}}. \bibinfo{pages}{1281--1282}.
\newblock


\bibitem[Fan et~al\mbox{.}(2024)]%
        {fan2024evaluating}
\bibfield{author}{\bibinfo{person}{Yimin Fan}, \bibinfo{person}{Fahim Dalvi}, \bibinfo{person}{Nadir Durrani}, {and} \bibinfo{person}{Hassan Sajjad}.} \bibinfo{year}{2024}\natexlab{}.
\newblock \showarticletitle{Evaluating neuron interpretation methods of nlp models}.
\newblock \bibinfo{journal}{\emph{Advances in Neural Information Processing Systems}}  \bibinfo{volume}{36} (\bibinfo{year}{2024}).
\newblock


\bibitem[Foote et~al\mbox{.}(2023)]%
        {foote2023neuron}
\bibfield{author}{\bibinfo{person}{Alex Foote}, \bibinfo{person}{Neel Nanda}, \bibinfo{person}{Esben Kran}, \bibinfo{person}{Ioannis Konstas}, \bibinfo{person}{Shay Cohen}, {and} \bibinfo{person}{Fazl Barez}.} \bibinfo{year}{2023}\natexlab{}.
\newblock \showarticletitle{Neuron to graph: Interpreting language model neurons at scale}.
\newblock \bibinfo{journal}{\emph{arXiv preprint arXiv:2305.19911}} (\bibinfo{year}{2023}).
\newblock


\bibitem[Gao et~al\mbox{.}(2024)]%
        {gao2024scaling}
\bibfield{author}{\bibinfo{person}{Leo Gao}, \bibinfo{person}{Tom~Dupr{\'e} la Tour}, \bibinfo{person}{Henk Tillman}, \bibinfo{person}{Gabriel Goh}, \bibinfo{person}{Rajan Troll}, \bibinfo{person}{Alec Radford}, \bibinfo{person}{Ilya Sutskever}, \bibinfo{person}{Jan Leike}, {and} \bibinfo{person}{Jeffrey Wu}.} \bibinfo{year}{2024}\natexlab{}.
\newblock \showarticletitle{Scaling and evaluating sparse autoencoders}.
\newblock \bibinfo{journal}{\emph{arXiv preprint arXiv:2406.04093}} (\bibinfo{year}{2024}).
\newblock


\bibitem[Gao et~al\mbox{.}(2023)]%
        {ChatRecGao2023}
\bibfield{author}{\bibinfo{person}{Yunfan Gao}, \bibinfo{person}{Tao Sheng}, \bibinfo{person}{Youlin Xiang}, \bibinfo{person}{Yun Xiong}, \bibinfo{person}{Haofen Wang}, {and} \bibinfo{person}{Jiawei Zhang}.} \bibinfo{year}{2023}\natexlab{}.
\newblock \bibinfo{title}{Chat-REC: Towards Interactive and Explainable LLMs-Augmented Recommender System}.
\newblock
\showeprint[arxiv]{2303.14524}~[cs.IR]
\urldef\tempurl%
\url{https://arxiv.org/abs/2303.14524}
\showURL{%
\tempurl}


\bibitem[Ge et~al\mbox{.}(2024)]%
        {ge2024survey}
\bibfield{author}{\bibinfo{person}{Yingqiang Ge}, \bibinfo{person}{Shuchang Liu}, \bibinfo{person}{Zuohui Fu}, \bibinfo{person}{Juntao Tan}, \bibinfo{person}{Zelong Li}, \bibinfo{person}{Shuyuan Xu}, \bibinfo{person}{Yunqi Li}, \bibinfo{person}{Yikun Xian}, {and} \bibinfo{person}{Yongfeng Zhang}.} \bibinfo{year}{2024}\natexlab{}.
\newblock \showarticletitle{A survey on trustworthy recommender systems}.
\newblock \bibinfo{journal}{\emph{ACM Transactions on Recommender Systems}} \bibinfo{volume}{3}, \bibinfo{number}{2} (\bibinfo{year}{2024}), \bibinfo{pages}{1--68}.
\newblock


\bibitem[Godoy and Amandi(2006)]%
        {godoy2006modeling}
\bibfield{author}{\bibinfo{person}{Daniela Godoy} {and} \bibinfo{person}{Anal{\'\i}a Amandi}.} \bibinfo{year}{2006}\natexlab{}.
\newblock \showarticletitle{Modeling user interests by conceptual clustering}.
\newblock \bibinfo{journal}{\emph{Information Systems}} \bibinfo{volume}{31}, \bibinfo{number}{4-5} (\bibinfo{year}{2006}), \bibinfo{pages}{247--265}.
\newblock


\bibitem[Hada et~al\mbox{.}(2021)]%
        {ReXPlugHada2021}
\bibfield{author}{\bibinfo{person}{Deepesh~V. Hada}, \bibinfo{person}{Vijaikumar M.}, {and} \bibinfo{person}{Shirish~K. Shevade}.} \bibinfo{year}{2021}\natexlab{}.
\newblock \showarticletitle{ReXPlug: Explainable Recommendation using Plug-and-Play Language Model}. In \bibinfo{booktitle}{\emph{Proceedings of the 44th International ACM SIGIR Conference on Research and Development in Information Retrieval}} (Virtual Event, Canada) \emph{(\bibinfo{series}{SIGIR '21})}. \bibinfo{publisher}{Association for Computing Machinery}, \bibinfo{address}{New York, NY, USA}, \bibinfo{pages}{81–91}.
\newblock
\showISBNx{9781450380379}
\href{https://doi.org/10.1145/3404835.3462939}{doi:\nolinkurl{10.1145/3404835.3462939}}


\bibitem[Harper and Konstan(2015)]%
        {harper2015movielens}
\bibfield{author}{\bibinfo{person}{F~Maxwell Harper} {and} \bibinfo{person}{Joseph~A Konstan}.} \bibinfo{year}{2015}\natexlab{}.
\newblock \showarticletitle{The movielens datasets: History and context}.
\newblock \bibinfo{journal}{\emph{Acm transactions on interactive intelligent systems (tiis)}} (\bibinfo{year}{2015}).
\newblock


\bibitem[He et~al\mbox{.}(2020)]%
        {he2020lightgcn}
\bibfield{author}{\bibinfo{person}{Xiangnan He}, \bibinfo{person}{Kuan Deng}, \bibinfo{person}{Xiang Wang}, \bibinfo{person}{Yan Li}, \bibinfo{person}{Yongdong Zhang}, {and} \bibinfo{person}{Meng Wang}.} \bibinfo{year}{2020}\natexlab{}.
\newblock \showarticletitle{Lightgcn: Simplifying and powering graph convolution network for recommendation}. In \bibinfo{booktitle}{\emph{Proceedings of the 43rd International ACM SIGIR conference on research and development in Information Retrieval}}. \bibinfo{pages}{639--648}.
\newblock


\bibitem[Jazayeri and Ostojic(2021)]%
        {jazayeri2021interpreting}
\bibfield{author}{\bibinfo{person}{Mehrdad Jazayeri} {and} \bibinfo{person}{Srdjan Ostojic}.} \bibinfo{year}{2021}\natexlab{}.
\newblock \showarticletitle{Interpreting neural computations by examining intrinsic and embedding dimensionality of neural activity}.
\newblock \bibinfo{journal}{\emph{Current opinion in neurobiology}}  \bibinfo{volume}{70} (\bibinfo{year}{2021}), \bibinfo{pages}{113--120}.
\newblock


\bibitem[Jermyn and Templeton(2024)]%
        {claude_ghost_grads}
\bibfield{author}{\bibinfo{person}{Adam Jermyn} {and} \bibinfo{person}{Adly Templeton}.} \bibinfo{year}{2024}\natexlab{}.
\newblock \showarticletitle{Ghost grads: An improvement on resampling}.
\newblock \bibinfo{journal}{\emph{Transformer Circuits Thread}} (\bibinfo{year}{2024}).
\newblock
\urldef\tempurl%
\url{https://transformer-circuits.pub/2024/jan-update/index.html#dict-learningresampling}
\showURL{%
\tempurl}


\bibitem[Kaffes et~al\mbox{.}(2021)]%
        {Kaffes2021Conterfactual}
\bibfield{author}{\bibinfo{person}{Vassilis Kaffes}, \bibinfo{person}{Dimitris Sacharidis}, {and} \bibinfo{person}{Giorgos Giannopoulos}.} \bibinfo{year}{2021}\natexlab{}.
\newblock \showarticletitle{Model-Agnostic Counterfactual Explanations of Recommendations}. In \bibinfo{booktitle}{\emph{Proceedings of the 29th ACM Conference on User Modeling, Adaptation and Personalization}} (Utrecht, Netherlands) \emph{(\bibinfo{series}{UMAP '21})}. \bibinfo{publisher}{Association for Computing Machinery}, \bibinfo{address}{New York, NY, USA}, \bibinfo{pages}{280–285}.
\newblock
\showISBNx{9781450383660}
\href{https://doi.org/10.1145/3450613.3456846}{doi:\nolinkurl{10.1145/3450613.3456846}}


\bibitem[Kang and McAuley(2018)]%
        {kang2018self}
\bibfield{author}{\bibinfo{person}{Wang-Cheng Kang} {and} \bibinfo{person}{Julian McAuley}.} \bibinfo{year}{2018}\natexlab{}.
\newblock \showarticletitle{Self-attentive sequential recommendation}. In \bibinfo{booktitle}{\emph{2018 IEEE international conference on data mining (ICDM)}}. IEEE.
\newblock


\bibitem[Ko et~al\mbox{.}(2022)]%
        {ko2022survey}
\bibfield{author}{\bibinfo{person}{Hyeyoung Ko}, \bibinfo{person}{Suyeon Lee}, \bibinfo{person}{Yoonseo Park}, {and} \bibinfo{person}{Anna Choi}.} \bibinfo{year}{2022}\natexlab{}.
\newblock \showarticletitle{A survey of recommendation systems: recommendation models, techniques, and application fields}.
\newblock \bibinfo{journal}{\emph{Electronics}} \bibinfo{volume}{11}, \bibinfo{number}{1} (\bibinfo{year}{2022}), \bibinfo{pages}{141}.
\newblock


\bibitem[Lei et~al\mbox{.}(2023)]%
        {lei2023recexplainer}
\bibfield{author}{\bibinfo{person}{Yuxuan Lei}, \bibinfo{person}{Jianxun Lian}, \bibinfo{person}{Jing Yao}, \bibinfo{person}{Xu Huang}, \bibinfo{person}{Defu Lian}, {and} \bibinfo{person}{Xing Xie}.} \bibinfo{year}{2023}\natexlab{}.
\newblock \showarticletitle{Recexplainer: Aligning large language models for recommendation model interpretability}.
\newblock \bibinfo{journal}{\emph{arXiv preprint arXiv:2311.10947}} (\bibinfo{year}{2023}).
\newblock


\bibitem[Leskovec et~al\mbox{.}(2007)]%
        {leskovec2007dynamics}
\bibfield{author}{\bibinfo{person}{Jure Leskovec}, \bibinfo{person}{Lada~A Adamic}, {and} \bibinfo{person}{Bernardo~A Huberman}.} \bibinfo{year}{2007}\natexlab{}.
\newblock \showarticletitle{The dynamics of viral marketing}.
\newblock \bibinfo{journal}{\emph{ACM Transactions on the Web (TWEB)}} \bibinfo{volume}{1}, \bibinfo{number}{1} (\bibinfo{year}{2007}), \bibinfo{pages}{5--es}.
\newblock


\bibitem[Li et~al\mbox{.}(2024)]%
        {li2024rechorus2}
\bibfield{author}{\bibinfo{person}{Jiayu Li}, \bibinfo{person}{Hanyu Li}, \bibinfo{person}{Zhiyu He}, \bibinfo{person}{Weizhi Ma}, \bibinfo{person}{Peijie Sun}, \bibinfo{person}{Min Zhang}, {and} \bibinfo{person}{Shaoping Ma}.} \bibinfo{year}{2024}\natexlab{}.
\newblock \showarticletitle{ReChorus2. 0: A Modular and Task-Flexible Recommendation Library}.
\newblock \bibinfo{journal}{\emph{arXiv preprint arXiv:2405.18058}} (\bibinfo{year}{2024}).
\newblock


\bibitem[Li et~al\mbox{.}(2021)]%
        {Li2021PETER}
\bibfield{author}{\bibinfo{person}{Lei Li}, \bibinfo{person}{Yongfeng Zhang}, {and} \bibinfo{person}{Li Chen}.} \bibinfo{year}{2021}\natexlab{}.
\newblock \bibinfo{title}{Personalized Transformer for Explainable Recommendation}.
\newblock
\showeprint[arxiv]{2105.11601}~[cs.IR]
\urldef\tempurl%
\url{https://arxiv.org/abs/2105.11601}
\showURL{%
\tempurl}


\bibitem[Li et~al\mbox{.}(2023)]%
        {Li2023PEPLER}
\bibfield{author}{\bibinfo{person}{Lei Li}, \bibinfo{person}{Yongfeng Zhang}, {and} \bibinfo{person}{Li Chen}.} \bibinfo{year}{2023}\natexlab{}.
\newblock \showarticletitle{Personalized prompt learning for explainable recommendation}.
\newblock \bibinfo{journal}{\emph{ACM Transactions on Information Systems}} (\bibinfo{year}{2023}).
\newblock


\bibitem[Liang et~al\mbox{.}(2021)]%
        {liang2021explaining}
\bibfield{author}{\bibinfo{person}{Yu Liang}, \bibinfo{person}{Siguang Li}, \bibinfo{person}{Chungang Yan}, \bibinfo{person}{Maozhen Li}, {and} \bibinfo{person}{Changjun Jiang}.} \bibinfo{year}{2021}\natexlab{}.
\newblock \showarticletitle{Explaining the black-box model: A survey of local interpretation methods for deep neural networks}.
\newblock \bibinfo{journal}{\emph{Neurocomputing}}  \bibinfo{volume}{419} (\bibinfo{year}{2021}), \bibinfo{pages}{168--182}.
\newblock


\bibitem[Liu et~al\mbox{.}(2019)]%
        {In2RecCIKM2019}
\bibfield{author}{\bibinfo{person}{Huafeng Liu}, \bibinfo{person}{Jingxuan Wen}, \bibinfo{person}{Liping Jing}, \bibinfo{person}{Jian Yu}, \bibinfo{person}{Xiangliang Zhang}, {and} \bibinfo{person}{Min Zhang}.} \bibinfo{year}{2019}\natexlab{}.
\newblock \showarticletitle{In2Rec: Influence-based Interpretable Recommendation}. In \bibinfo{booktitle}{\emph{Proceedings of the 28th ACM International Conference on Information and Knowledge Management}} (Beijing, China) \emph{(\bibinfo{series}{CIKM '19})}. \bibinfo{publisher}{Association for Computing Machinery}, \bibinfo{address}{New York, NY, USA}, \bibinfo{pages}{1803–1812}.
\newblock
\showISBNx{9781450369763}
\href{https://doi.org/10.1145/3357384.3358017}{doi:\nolinkurl{10.1145/3357384.3358017}}


\bibitem[Montavon et~al\mbox{.}(2018)]%
        {montavon2018methods}
\bibfield{author}{\bibinfo{person}{Gr{\'e}goire Montavon}, \bibinfo{person}{Wojciech Samek}, {and} \bibinfo{person}{Klaus-Robert M{\"u}ller}.} \bibinfo{year}{2018}\natexlab{}.
\newblock \showarticletitle{Methods for interpreting and understanding deep neural networks}.
\newblock \bibinfo{journal}{\emph{Digital signal processing}}  \bibinfo{volume}{73} (\bibinfo{year}{2018}), \bibinfo{pages}{1--15}.
\newblock


\bibitem[Nilashi et~al\mbox{.}(2016)]%
        {nilashi2016recommendation}
\bibfield{author}{\bibinfo{person}{Mehrbakhsh Nilashi}, \bibinfo{person}{Dietmar Jannach}, \bibinfo{person}{Othman bin Ibrahim}, \bibinfo{person}{Mohammad~Dalvi Esfahani}, {and} \bibinfo{person}{Hossein Ahmadi}.} \bibinfo{year}{2016}\natexlab{}.
\newblock \showarticletitle{Recommendation quality, transparency, and website quality for trust-building in recommendation agents}.
\newblock \bibinfo{journal}{\emph{Electronic commerce research and applications}}  \bibinfo{volume}{19} (\bibinfo{year}{2016}), \bibinfo{pages}{70--84}.
\newblock


\bibitem[Ning and Karypis(2011)]%
        {ning2011slim}
\bibfield{author}{\bibinfo{person}{Xia Ning} {and} \bibinfo{person}{George Karypis}.} \bibinfo{year}{2011}\natexlab{}.
\newblock \showarticletitle{Slim: Sparse linear methods for top-n recommender systems}. In \bibinfo{booktitle}{\emph{2011 IEEE 11th international conference on data mining}}. IEEE, \bibinfo{pages}{497--506}.
\newblock


\bibitem[Patel et~al\mbox{.}(2023)]%
        {patel2023recommendation}
\bibfield{author}{\bibinfo{person}{Dhruval Patel}, \bibinfo{person}{Foram Patel}, {and} \bibinfo{person}{Uttam Chauhan}.} \bibinfo{year}{2023}\natexlab{}.
\newblock \showarticletitle{Recommendation systems: Types, applications, and challenges}.
\newblock \bibinfo{journal}{\emph{International Journal of Computing and Digital Systems}} (\bibinfo{year}{2023}).
\newblock


\bibitem[Rajput et~al\mbox{.}(2023)]%
        {rajput2023recommender}
\bibfield{author}{\bibinfo{person}{Shashank Rajput}, \bibinfo{person}{Nikhil Mehta}, \bibinfo{person}{Anima Singh}, \bibinfo{person}{Raghunandan Hulikal~Keshavan}, \bibinfo{person}{Trung Vu}, \bibinfo{person}{Lukasz Heldt}, \bibinfo{person}{Lichan Hong}, \bibinfo{person}{Yi Tay}, \bibinfo{person}{Vinh Tran}, \bibinfo{person}{Jonah Samost}, {et~al\mbox{.}}} \bibinfo{year}{2023}\natexlab{}.
\newblock \showarticletitle{Recommender systems with generative retrieval}.
\newblock \bibinfo{journal}{\emph{Advances in Neural Information Processing Systems}}  \bibinfo{volume}{36} (\bibinfo{year}{2023}), \bibinfo{pages}{10299--10315}.
\newblock


\bibitem[Rendle et~al\mbox{.}(2012)]%
        {rendle2012bpr}
\bibfield{author}{\bibinfo{person}{Steffen Rendle}, \bibinfo{person}{Christoph Freudenthaler}, \bibinfo{person}{Zeno Gantner}, {and} \bibinfo{person}{Lars Schmidt-Thieme}.} \bibinfo{year}{2012}\natexlab{}.
\newblock \showarticletitle{BPR: Bayesian personalized ranking from implicit feedback}.
\newblock \bibinfo{journal}{\emph{arXiv preprint arXiv:1205.2618}} (\bibinfo{year}{2012}).
\newblock


\bibitem[Rousseeuw(1987)]%
        {rousseeuw1987silhouettes}
\bibfield{author}{\bibinfo{person}{Peter~J Rousseeuw}.} \bibinfo{year}{1987}\natexlab{}.
\newblock \showarticletitle{Silhouettes: a graphical aid to the interpretation and validation of cluster analysis}.
\newblock \bibinfo{journal}{\emph{Journal of computational and applied mathematics}}  \bibinfo{volume}{20} (\bibinfo{year}{1987}), \bibinfo{pages}{53--65}.
\newblock


\bibitem[Sajjad et~al\mbox{.}(2022)]%
        {sajjad2022neuron}
\bibfield{author}{\bibinfo{person}{Hassan Sajjad}, \bibinfo{person}{Nadir Durrani}, {and} \bibinfo{person}{Fahim Dalvi}.} \bibinfo{year}{2022}\natexlab{}.
\newblock \showarticletitle{Neuron-level interpretation of deep nlp models: A survey}.
\newblock \bibinfo{journal}{\emph{Transactions of the Association for Computational Linguistics}}  \bibinfo{volume}{10} (\bibinfo{year}{2022}), \bibinfo{pages}{1285--1303}.
\newblock


\bibitem[Schedl et~al\mbox{.}(2022)]%
        {schedl2022lfm}
\bibfield{author}{\bibinfo{person}{Markus Schedl}, \bibinfo{person}{Stefan Brandl}, \bibinfo{person}{Oleg Lesota}, \bibinfo{person}{Emilia Parada-Cabaleiro}, \bibinfo{person}{David Penz}, {and} \bibinfo{person}{Navid Rekabsaz}.} \bibinfo{year}{2022}\natexlab{}.
\newblock \showarticletitle{LFM-2b: A dataset of enriched music listening events for recommender systems research and fairness analysis}. In \bibinfo{booktitle}{\emph{Proceedings of the 2022 Conference on Human Information Interaction and Retrieval}}. \bibinfo{pages}{337--341}.
\newblock


\bibitem[Sethi et~al\mbox{.}(2022)]%
        {sethi2022recshard}
\bibfield{author}{\bibinfo{person}{Geet Sethi}, \bibinfo{person}{Bilge Acun}, \bibinfo{person}{Niket Agarwal}, \bibinfo{person}{Christos Kozyrakis}, \bibinfo{person}{Caroline Trippel}, {and} \bibinfo{person}{Carole-Jean Wu}.} \bibinfo{year}{2022}\natexlab{}.
\newblock \showarticletitle{RecShard: statistical feature-based memory optimization for industry-scale neural recommendation}. In \bibinfo{booktitle}{\emph{Proceedings of the 27th ACM International Conference on Architectural Support for Programming Languages and Operating Systems}}. \bibinfo{pages}{344--358}.
\newblock


\bibitem[Sinha and Swearingen(2002)]%
        {sinha2002role}
\bibfield{author}{\bibinfo{person}{Rashmi Sinha} {and} \bibinfo{person}{Kirsten Swearingen}.} \bibinfo{year}{2002}\natexlab{}.
\newblock \showarticletitle{The role of transparency in recommender systems}. In \bibinfo{booktitle}{\emph{CHI'02 extended abstracts on Human factors in computing systems}}. \bibinfo{pages}{830--831}.
\newblock


\bibitem[Steck(2019)]%
        {steck2019embarrassingly}
\bibfield{author}{\bibinfo{person}{Harald Steck}.} \bibinfo{year}{2019}\natexlab{}.
\newblock \showarticletitle{Embarrassingly shallow autoencoders for sparse data}. In \bibinfo{booktitle}{\emph{The World Wide Web Conference}}. \bibinfo{pages}{3251--3257}.
\newblock


\bibitem[Tal et~al\mbox{.}(2021)]%
        {Tal21Attention}
\bibfield{author}{\bibinfo{person}{Omer Tal}, \bibinfo{person}{Yang Liu}, \bibinfo{person}{Jimmy Huang}, \bibinfo{person}{Xiaohui Yu}, {and} \bibinfo{person}{Bushra Aljbawi}.} \bibinfo{year}{2021}\natexlab{}.
\newblock \showarticletitle{Neural Attention Frameworks for Explainable Recommendation}.
\newblock \bibinfo{journal}{\emph{IEEE Transactions on Knowledge and Data Engineering}} \bibinfo{volume}{33}, \bibinfo{number}{5} (\bibinfo{year}{2021}), \bibinfo{pages}{2137--2150}.
\newblock


\bibitem[Tan et~al\mbox{.}(2021)]%
        {Tan2021Counterfactual}
\bibfield{author}{\bibinfo{person}{Juntao Tan}, \bibinfo{person}{Shuyuan Xu}, \bibinfo{person}{Yingqiang Ge}, \bibinfo{person}{Yunqi Li}, \bibinfo{person}{Xu Chen}, {and} \bibinfo{person}{Yongfeng Zhang}.} \bibinfo{year}{2021}\natexlab{}.
\newblock \showarticletitle{Counterfactual Explainable Recommendation}. In \bibinfo{booktitle}{\emph{Proceedings of the 30th ACM International Conference on Information and Knowledge Management}}. \bibinfo{publisher}{ACM}, \bibinfo{pages}{1784–1793}.
\newblock
\href{https://doi.org/10.1145/3459637.3482420}{doi:\nolinkurl{10.1145/3459637.3482420}}


\bibitem[Templeton et~al\mbox{.}(2024)]%
        {claude_scaling}
\bibfield{author}{\bibinfo{person}{Adly Templeton}, \bibinfo{person}{Tom Conerly}, \bibinfo{person}{Jonathan Marcus}, {and} \bibinfo{person}{Others}.} \bibinfo{year}{2024}\natexlab{}.
\newblock \showarticletitle{Scaling Monosemanticity: Extracting Interpretable Features from Claude 3 Sonnet}.
\newblock \bibinfo{journal}{\emph{Anthropic}} (\bibinfo{year}{2024}).
\newblock


\bibitem[To{\v{s}}i{\'c} and Frossard(2011)]%
        {tovsic2011dictionary}
\bibfield{author}{\bibinfo{person}{Ivana To{\v{s}}i{\'c}} {and} \bibinfo{person}{Pascal Frossard}.} \bibinfo{year}{2011}\natexlab{}.
\newblock \showarticletitle{Dictionary learning}.
\newblock \bibinfo{journal}{\emph{IEEE Signal Processing Magazine}} \bibinfo{volume}{28}, \bibinfo{number}{2} (\bibinfo{year}{2011}), \bibinfo{pages}{27--38}.
\newblock


\bibitem[Tsang et~al\mbox{.}(2020)]%
        {featureinteractiontsang2020}
\bibfield{author}{\bibinfo{person}{Michael Tsang}, \bibinfo{person}{Dehua Cheng}, \bibinfo{person}{Hanpeng Liu}, \bibinfo{person}{Xue Feng}, \bibinfo{person}{Eric Zhou}, {and} \bibinfo{person}{Yan Liu}.} \bibinfo{year}{2020}\natexlab{}.
\newblock \bibinfo{title}{Feature Interaction Interpretability: A Case for Explaining Ad-Recommendation Systems via Neural Interaction Detection}.
\newblock
\showeprint[arxiv]{2006.10966}~[stat.ML]
\urldef\tempurl%
\url{https://arxiv.org/abs/2006.10966}
\showURL{%
\tempurl}


\bibitem[Wang et~al\mbox{.}(2022)]%
        {wang2022target}
\bibfield{author}{\bibinfo{person}{Chenyang Wang}, \bibinfo{person}{Zhefan Wang}, \bibinfo{person}{Yankai Liu}, \bibinfo{person}{Yang Ge}, \bibinfo{person}{Weizhi Ma}, \bibinfo{person}{Min Zhang}, \bibinfo{person}{Yiqun Liu}, \bibinfo{person}{Junlan Feng}, \bibinfo{person}{Chao Deng}, {and} \bibinfo{person}{Shaoping Ma}.} \bibinfo{year}{2022}\natexlab{}.
\newblock \showarticletitle{Target interest distillation for multi-interest recommendation}. In \bibinfo{booktitle}{\emph{Proceedings of the 31st ACM International Conference on Information \& Knowledge Management}}.
\newblock


\bibitem[Wang et~al\mbox{.}(2024)]%
        {wang2024learnable}
\bibfield{author}{\bibinfo{person}{Wenjie Wang}, \bibinfo{person}{Honghui Bao}, \bibinfo{person}{Xinyu Lin}, \bibinfo{person}{Jizhi Zhang}, \bibinfo{person}{Yongqi Li}, \bibinfo{person}{Fuli Feng}, \bibinfo{person}{See-Kiong Ng}, {and} \bibinfo{person}{Tat-Seng Chua}.} \bibinfo{year}{2024}\natexlab{}.
\newblock \showarticletitle{Learnable item tokenization for generative recommendation}. In \bibinfo{booktitle}{\emph{Proceedings of the 33rd ACM International Conference on Information and Knowledge Management}}. \bibinfo{pages}{2400--2409}.
\newblock


\bibitem[Wang et~al\mbox{.}(2018a)]%
        {RLWang2018}
\bibfield{author}{\bibinfo{person}{Xiting Wang}, \bibinfo{person}{Yiru Chen}, \bibinfo{person}{Jie Yang}, \bibinfo{person}{Le Wu}, \bibinfo{person}{Zhengtao Wu}, {and} \bibinfo{person}{Xing Xie}.} \bibinfo{year}{2018}\natexlab{a}.
\newblock \showarticletitle{A Reinforcement Learning Framework for Explainable Recommendation}. In \bibinfo{booktitle}{\emph{2018 IEEE International Conference on Data Mining (ICDM)}}. \bibinfo{pages}{587--596}.
\newblock
\href{https://doi.org/10.1109/ICDM.2018.00074}{doi:\nolinkurl{10.1109/ICDM.2018.00074}}


\bibitem[Wang et~al\mbox{.}(2018b)]%
        {Wang2018KPRN}
\bibfield{author}{\bibinfo{person}{Xiang Wang}, \bibinfo{person}{Dingxian Wang}, \bibinfo{person}{Canran Xu}, \bibinfo{person}{Xiangnan He}, \bibinfo{person}{Yixin Cao}, {and} \bibinfo{person}{Tat-Seng Chua}.} \bibinfo{year}{2018}\natexlab{b}.
\newblock \bibinfo{title}{Explainable Reasoning over Knowledge Graphs for Recommendation}.
\newblock
\showeprint[arxiv]{1811.04540}~[cs.IR]
\urldef\tempurl%
\url{https://arxiv.org/abs/1811.04540}
\showURL{%
\tempurl}


\bibitem[Xian et~al\mbox{.}(2020)]%
        {Xian2020CAFE}
\bibfield{author}{\bibinfo{person}{Yikun Xian}, \bibinfo{person}{Zuohui Fu}, \bibinfo{person}{Handong Zhao}, \bibinfo{person}{Yingqiang Ge}, \bibinfo{person}{Xu Chen}, \bibinfo{person}{Qiaoying Huang}, \bibinfo{person}{Shijie Geng}, \bibinfo{person}{Zhou Qin}, \bibinfo{person}{Gerard de Melo}, \bibinfo{person}{S. Muthukrishnan}, {and} \bibinfo{person}{Yongfeng Zhang}.} \bibinfo{year}{2020}\natexlab{}.
\newblock \showarticletitle{CAFE: Coarse-to-Fine Neural Symbolic Reasoning for Explainable Recommendation}. In \bibinfo{booktitle}{\emph{Proceedings of the 29th ACM International Conference on Information and Knowledge Management}}. \bibinfo{publisher}{ACM}, \bibinfo{pages}{1645–1654}.
\newblock
\href{https://doi.org/10.1145/3340531.3412038}{doi:\nolinkurl{10.1145/3340531.3412038}}


\bibitem[Xiao et~al\mbox{.}(2017)]%
        {Xiao2017AFM}
\bibfield{author}{\bibinfo{person}{Jun Xiao}, \bibinfo{person}{Hao Ye}, \bibinfo{person}{Xiangnan He}, \bibinfo{person}{Hanwang Zhang}, \bibinfo{person}{Fei Wu}, {and} \bibinfo{person}{Tat-Seng Chua}.} \bibinfo{year}{2017}\natexlab{}.
\newblock \bibinfo{title}{Attentional Factorization Machines: Learning the Weight of Feature Interactions via Attention Networks}.
\newblock
\urldef\tempurl%
\url{https://arxiv.org/abs/1708.04617}
\showURL{%
\tempurl}


\bibitem[Xu et~al\mbox{.}(2023)]%
        {Xu2023Distillation}
\bibfield{author}{\bibinfo{person}{Zhichao Xu}, \bibinfo{person}{Hansi Zeng}, \bibinfo{person}{Juntao Tan}, \bibinfo{person}{Zuohui Fu}, \bibinfo{person}{Yongfeng Zhang}, {and} \bibinfo{person}{Qingyao Ai}.} \bibinfo{year}{2023}\natexlab{}.
\newblock \showarticletitle{A Reusable Model-agnostic Framework for Faithfully Explainable Recommendation and System Scrutability}.
\newblock \bibinfo{journal}{\emph{ACM Trans. Inf. Syst.}} \bibinfo{volume}{42}, \bibinfo{number}{1}, Article \bibinfo{articleno}{29} (\bibinfo{date}{Aug.} \bibinfo{year}{2023}), \bibinfo{numpages}{29}~pages.
\newblock
\showISSN{1046-8188}
\href{https://doi.org/10.1145/3605357}{doi:\nolinkurl{10.1145/3605357}}


\bibitem[Yelp(2018)]%
        {Yelp2018}
\bibfield{author}{\bibinfo{person}{Yelp}.} \bibinfo{year}{2018}\natexlab{}.
\newblock \showarticletitle{Yelp Open Dataset}.
\newblock
\urldef\tempurl%
\url{https://business.yelp.com/data/resources/open-dataset/}
\showURL{%
\tempurl}


\bibitem[Zhang et~al\mbox{.}(2024)]%
        {FENCRZhang2024}
\bibfield{author}{\bibinfo{person}{Xiaoyu Zhang}, \bibinfo{person}{Shaoyun Shi}, \bibinfo{person}{Yishan Li}, \bibinfo{person}{Weizhi Ma}, \bibinfo{person}{Peijie Sun}, {and} \bibinfo{person}{Min Zhang}.} \bibinfo{year}{2024}\natexlab{}.
\newblock \showarticletitle{Feature-Enhanced Neural Collaborative Reasoning for Explainable Recommendation}.
\newblock \bibinfo{journal}{\emph{ACM Trans. Inf. Syst.}} (\bibinfo{date}{Aug.} \bibinfo{year}{2024}).
\newblock
\showISSN{1046-8188}
\href{https://doi.org/10.1145/3690381}{doi:\nolinkurl{10.1145/3690381}}


\bibitem[Zhang et~al\mbox{.}(2020a)]%
        {zhang2020explainable}
\bibfield{author}{\bibinfo{person}{Yongfeng Zhang}, \bibinfo{person}{Xu Chen}, {et~al\mbox{.}}} \bibinfo{year}{2020}\natexlab{a}.
\newblock \showarticletitle{Explainable recommendation: A survey and new perspectives}.
\newblock \bibinfo{journal}{\emph{Foundations and Trends{\textregistered} in Information Retrieval}} \bibinfo{volume}{14}, \bibinfo{number}{1} (\bibinfo{year}{2020}), \bibinfo{pages}{1--101}.
\newblock


\bibitem[Zhang et~al\mbox{.}(2014)]%
        {zhang2014explicit}
\bibfield{author}{\bibinfo{person}{Yongfeng Zhang}, \bibinfo{person}{Guokun Lai}, \bibinfo{person}{Min Zhang}, \bibinfo{person}{Yi Zhang}, \bibinfo{person}{Yiqun Liu}, {and} \bibinfo{person}{Shaoping Ma}.} \bibinfo{year}{2014}\natexlab{}.
\newblock \showarticletitle{Explicit factor models for explainable recommendation based on phrase-level sentiment analysis}. In \bibinfo{booktitle}{\emph{Proceedings of the 37th international ACM SIGIR conference on Research \& development in information retrieval}}. \bibinfo{pages}{83--92}.
\newblock


\bibitem[Zhang et~al\mbox{.}(2021)]%
        {zhang2021survey}
\bibfield{author}{\bibinfo{person}{Yu Zhang}, \bibinfo{person}{Peter Ti{\v{n}}o}, \bibinfo{person}{Ale{\v{s}} Leonardis}, {and} \bibinfo{person}{Ke Tang}.} \bibinfo{year}{2021}\natexlab{}.
\newblock \showarticletitle{A survey on neural network interpretability}.
\newblock \bibinfo{journal}{\emph{IEEE Transactions on Emerging Topics in Computational Intelligence}} \bibinfo{volume}{5}, \bibinfo{number}{5} (\bibinfo{year}{2021}), \bibinfo{pages}{726--742}.
\newblock


\bibitem[Zhang et~al\mbox{.}(2020b)]%
        {DistillationZhang2020}
\bibfield{author}{\bibinfo{person}{Yuan Zhang}, \bibinfo{person}{Xiaoran Xu}, \bibinfo{person}{Hanning Zhou}, {and} \bibinfo{person}{Yan Zhang}.} \bibinfo{year}{2020}\natexlab{b}.
\newblock \showarticletitle{Distilling Structured Knowledge into Embeddings for Explainable and Accurate Recommendation}. In \bibinfo{booktitle}{\emph{Proceedings of the 13th International Conference on Web Search and Data Mining}} (Houston, TX, USA) \emph{(\bibinfo{series}{WSDM '20})}. \bibinfo{publisher}{Association for Computing Machinery}, \bibinfo{address}{New York, NY, USA}, \bibinfo{pages}{735–743}.
\newblock
\showISBNx{9781450368223}
\href{https://doi.org/10.1145/3336191.3371790}{doi:\nolinkurl{10.1145/3336191.3371790}}


\bibitem[Zhao et~al\mbox{.}(2020)]%
        {KGZhao2020}
\bibfield{author}{\bibinfo{person}{Kangzhi Zhao}, \bibinfo{person}{Xiting Wang}, \bibinfo{person}{Yuren Zhang}, \bibinfo{person}{Li Zhao}, \bibinfo{person}{Zheng Liu}, \bibinfo{person}{Chunxiao Xing}, {and} \bibinfo{person}{Xing Xie}.} \bibinfo{year}{2020}\natexlab{}.
\newblock \showarticletitle{Leveraging Demonstrations for Reinforcement Recommendation Reasoning over Knowledge Graphs}. In \bibinfo{booktitle}{\emph{Proceedings of the 43rd International ACM SIGIR Conference on Research and Development in Information Retrieval}} (Virtual Event, China) \emph{(\bibinfo{series}{SIGIR '20})}. \bibinfo{publisher}{Association for Computing Machinery}, \bibinfo{address}{New York, NY, USA}, \bibinfo{pages}{239–248}.
\newblock
\showISBNx{9781450380164}
\href{https://doi.org/10.1145/3397271.3401171}{doi:\nolinkurl{10.1145/3397271.3401171}}


\end{thebibliography}

\appendix









\end{document}